\documentclass[twocolumn,superscriptaddress,longbibliography,prc,nofootinbib]{revtex4-2}
\usepackage{times}
\usepackage[T1]{fontenc}

\usepackage{physics}
\usepackage{amsmath,amssymb,mathrsfs}
\usepackage{graphicx}
\usepackage{dcolumn}
\usepackage{bm}
\usepackage{morefloats}
\usepackage{multirow}
\usepackage{amssymb}
\usepackage{amsmath}
\usepackage{mathtools}
\usepackage{xcolor}
\usepackage{longtable}
\usepackage{fix-cm}
\usepackage{verbatim}
\usepackage{float}
\usepackage{siunitx}
\usepackage{mathptmx} 
\usepackage[T1]{fontenc}
\usepackage{subfigure}
\usepackage[colorlinks,allcolors=blue]{hyperref}

\makeatletter
\def\NAT@def@citea{\def\@citea{\NAT@separator}}
\makeatother

\def\0\\{\nonumber\\}

\newcommand{\beq}{\begin{equation}}
\newcommand{\eeq}{\end{equation}}
\newcommand{\beqn}{\begin{eqnarray}}
\newcommand{\eeqn}{\end{eqnarray}}

\newcommand{\kfx}{k_{\mathrm{Fx}}}

\makeatletter
\newcommand\footnoteref[1]{\protected@xdef\@thefnmark{\ref{#1}}\@footnotemark}
\makeatother

\begin{document}


\title{Vortex creep heating in neutron star cooling with direct Urca processes in heavy neutron stars
}

\author{Yoonhak Nam}
\email{nam.y.b76c@m.isct.ac.jp}
\affiliation{Department of Physics, School of Science, Institute of Science Tokyo, Tokyo 152-8550, Japan}

\author{Kazuyuki Sekizawa}
\email{sekizawa@phys.sci.isct.ac.jp}
\affiliation{Department of Physics, School of Science, Institute of Science Tokyo, Tokyo 152-8550, Japan}
\affiliation{Nuclear Physics Division, Center for Computational Sciences, University of Tsukuba, Ibaraki 305-8577, Japan}
\affiliation{RIKEN Nishina Center, Saitama 351-0198, Japan}

\date{November 17, 2025}

\begin{abstract}
\edef\oldrightskip{\the\rightskip}
\begin{description}
\rightskip\oldrightskip\relax
\setlength{\parskip}{0pt} 

\item[Background]
Old, thermally bright neutron stars indicate internal heating operating at late times. Among proposed mechanisms, vortex creep heating (VCH) offers a robust link between rotational energy loss and frictional dissipation in the pinned inner-crust superfluid. Yet most VCH studies have focused on lighter stars or $t\!\gtrsim\!10^5$\,yr, leaving the interplay with fast direct Urca (DUrca) cooling in massive stars largely unexplored.

\item[Purpose]
We aim to (i) implement VCH in our cooling framework and validate it against prior results; (ii) establish a \emph{physically consistent} domain where the steady-state heating luminosity $L_{\mathrm{h}}=J|\dot{\Omega}_\infty|$ is applicable; (iii) quantify how $(B,P_0)$ regulate observable VCH signatures in the presence of DUrca cooling; and (iv) introduce a 3D representation $(\log_{10}t,\log_{10}T_{\mathrm{s}}^\infty,\log_{10}B)$ that resolves 
degeneracies inherent in two-dimensional cooling projections.

\item[Methods]
We compute one-dimensional general-relativistic cooling with BSk24 and APR equations of state, standard baryon pairing gap sets, and iron/carbon envelopes. VCH is modeled as an inner-crust heating source with luminosity 
$L_{\mathrm{h}} = J|\dot{\Omega}_\infty|$, where $J \simeq 10^{42.9\text{--}43.8}\,$erg\,s follows the steady-state VCH prescription. We introduce a \emph{quantum-creep coverage fraction} $f_{\mathrm{Q}}(t)$ to diagnose entry into the quantum tunneling regime and require $t_{\mathrm{sep}}\!>\!t_{\mathrm{Q}}$ (heating becomes visible only after quantum-creep is established) for computational validity. We survey $B\!=\!10^{10\text{--}13}$\,G and $P_0\!=\!10$--$570$\,ms for $1.4$ and $2.0\,M_\odot$, and compare with a curated set of ordinary pulsars with measured $(P,\dot P)$.

\item[Results]
(1) Our implementation reproduces published VCH bands; the superfluid-fraction switch $\chi_{\mathrm{sf}}$ changes $T_{\mathrm{s}}^\infty$ by $\lesssim0.7\%$ once heating matters.
(2) The validity boundary in $(B,P_0)$ follows the spin-down scaling expected from magnetic-dipole braking, demonstrating consistency with the $|\dot\Omega|$ evolution.
(3) DUrca+VCH tracks in $2.0\,M_\odot$ can maintain $T_{\mathrm{s}}^\infty\!\gtrsim\!10^5$\,K for $B\!\gtrsim\!10^{11\text{--}12}$\,G up to $P_0\!\sim\!\mathcal{O}(10^2)$\,ms, whereas cooling-only models overcool.
(4) The 3D plots resolve two-dimensional degeneracies by showing that sources that appear coincident in $(t, T_{\mathrm{s}}^{\infty})$ can in fact occupy distinct magnetic-field layers in $(t, T_{\mathrm{s}}^{\infty}, B)$, so similar-looking 2D tracks need not to correspond to a single evolutionary pathway.

\item[Conclusions]
VCH can substantially reshape late-time thermal states when spin-down power remains large; its observability depends far more on $(B,P_0)$ than on mass alone. We provide a practical $(B,P_0)$ validity map for using $L_{\mathrm{h}}=J|\dot{\Omega}_\infty|$ and advocate treating $B$ as a \emph{co-equal} axis in cooling analyses. The 3D representation mitigates interpretational ambiguities in population studies and sets the stage for upgrades to anisotropic conduction/field evolution, microscopic calibration of $J$ from pinning theory, and data-driven exploration of $(P\dot P, P_0, M, \Delta M)$ against current observational samples.

\end{description}
\end{abstract}

\maketitle

\section{Introduction}\label{Sec:Intro}

Neutron stars are compact remnants left behind after core-collapse supernovae. Their central densities exceed nuclear saturation density ($\rho_0 \simeq 2.7\times10^{14}\ \mathrm{g\,cm^{-3}},\ n_0 \simeq 0.16\ \mathrm{fm^{-3}}$), thereby probing regimes of dense nuclear matter unattainable in terrestrial laboratories. As such, neutron stars serve as unique natural laboratories for studying matter under extreme conditions. However, even the nearest known neutron star, RX~J1856.5$-$3754, lies at a distance of $\sim 400$ light-years \cite{Walter_2010_RXJ18575-3754_dist}, making direct exploration impossible with current technology. Consequently, our understanding of neutron star interiors must be inferred indirectly through observable properties such as rotation period $P$, spindown rate $\dot{P}$, mass, radius, surface temperature, luminosity, magnetic field, electromagnetic spectra, and pulse profiles.

In this study, we focus on neutron star cooling---the thermal evolution traced by surface temperature and luminosity---as a key observable of dense-matter interiors of neutron stars. Because thermal evolution depends sensitively on the internal composition and structure, theoretical cooling curves can be compared with observations to probe the microphysics of the core and crust (\textit{e.g.}, Refs.~\cite{Yakovlev_2004_neutron_star_cooling, Page_2006_the_cooling_of_compact_stars}). Early theoretical works established that neutrino emission governs the cooling of young neutron stars, while photon radiation dominates at late times. Within the neutrino-cooling era, the standard channels are the modified Urca process and nucleon-nucleon bremsstrahlung (\textit{e.g.}, Refs.~\cite{Friman_1979_murca_brems, Yakovlev_2001_neutrino_emission_165pages}), whereas enhanced cooling via direct Urca (DUrca) becomes possible in sufficiently massive stars once composition thresholds are met \cite{Lattimer_1991_durca}. The surface photon emissions and the relation between interior and surface temperatures depend strongly on the heat-blanketing envelope and magnetic field (\textit{e.g.}, Refs.~\cite{Gudmundsson_1983_envelope, Potekhin_1997_accreted_envelope, Potekhin_2015_review_cooling}).

A major breakthrough came with \textit{Chandra}’s real-time monitoring of the Cassiopeia~A neutron star (Cas~A~NS), whose observed decade-long temperature decline \cite{Heinke_2010_cas_a_fast_cooling} was most naturally interpreted as the onset of neutron $^3\mathrm{P}_2$ pairing and the associated Cooper-pair breaking and formation (PBF) neutrino emission \cite{Page_2011_cas_a_pbf_process, shternin_2011_nt_SYHHP}, going beyond the predictions of standard modified Urca and bremsstrahlung cooling.

At the same time, a growing population of old neutron stars has been observed (\textit{e.g.}, Refs.~\cite{Becker_2004_heating_evidence_1, Zavlin_2004_heating_evidence_2, Misanovic_2008_heating_evidence_3}) with surface temperatures higher than expected from cooling alone, indicating the existence of heating \cite{Gonzalez_2010_internal_heating}. Various mechanisms have been proposed, including rotochemical heating (out-of-equilibrium beta reactions driven by spindown compression) \cite{Reisenegger_1995_rotochemical_1, Fernandez_2005_rotochemical_2, Petrovich_2010_rotochemical_3}, magnetic field decay \cite{Pons_2009_magnetic_field_decay_1, Vigano_2013_magnetic_field_decay_2}, dark-matter heating \cite{Kouvaris_2008_dark_matter_1, Lavallaz_2010_dark_matter_2, Baryakhtar_2017_dark_matter_3}, and vortex creep heating \cite{Alpar_1984_VCH_1, Shibazaki_1989_VCH_2, Link_1996_VCH_3}.

Among these, vortex creep heating (VCH) offers a particularly compelling and physically well-grounded explanation for the thermal emission of old, isolated neutron stars. In the rotating inner crust, neutron superfluid vortices carry quantized circulation and interact with the nuclear lattice, leading to pinning. As the neutron star spins down, which co-rotate with the normal component (nuclear lattice) due to strong magnetic field, a lag $\delta\Omega$ develops between the superfluid and the crust; the resulting Magnus force drives vortices outward while pinning resists motion. Through thermal activation or quantum tunneling, vortices episodically unpin and “creep,” dissipating rotational energy as heat \cite{Alpar_1984_VCH_1, Shibazaki_1989_VCH_2, Link_1996_VCH_3}. The critical lag is approximately given by
\begin{align}
\delta\Omega_{\rm cr}(r) \simeq \frac{f_{\rm pin}(r)}{n_\text{s}(r)\,\kappa\,r},
\end{align}
where $f_{\rm pin}$ is the pinning force per unit length, $n_\text{s}$ the superfluid density, $r$ the cylindrical radius, and $\kappa=h/(2m_\text{n})$ the quantum of circulation. In the late-time, low-temperature regime relevant for $t\gtrsim10^5\ \mathrm{yr}$, quantum creep---vortex creep mediated by quantum tunneling---dominates, maintaining a steady state with $\delta\Omega_{\infty}\approx\delta\Omega_{\rm cr}$. The corresponding frictional power ( \textit{i.e.}, heating luminosity) $L_\mathrm{h}$ scales with the currently observed spin-down rate $\dot{\Omega}_\infty$ as
\begin{align}
L_{\mathrm{h}} &= J\,|\dot{\Omega}_\infty|,\\
J &\equiv \int_{\rm pin} \,\delta\Omega_{\infty}(r)\, \dd I_{\mathrm{p}},
\end{align}
where the proportional constant $J$ encapsulates the dependence on stellar structure and microscopic pinning physics 
\cite{Donati_2004_pinning_physics_1, Seveso_2015_pinning_physics_2, Fujiwara_2024}. 
Since $J$ has an approximately universal value for all neutron stars, the late-time thermal state can be expressed as a function of $J$ and $|\dot{\Omega}_\infty|$ \cite{Fujiwara_2024}.

Recent studies have demonstrated that such a creep-powered luminosity reproduces the elevated temperatures of old neutron stars with $t>10^5\ \mathrm{yr}$ \cite{Gonzalez_2010_internal_heating, Fujiwara_2024}. In particular, under low and isothermal internal temperatures typical of these ages, quantum creep establishes a stable steady state with $L_{\mathrm{h}}=J|\dot{\Omega}|$, explaining multiple warm old neutron stars \cite{Fujiwara_2024}.

The potential importance of VCH in \emph{massive} neutron stars is especially noteworthy. High-mass stars likely undergo rapid early cooling via DUrca, driving them to low temperatures at young ages \cite{Lattimer_1991_durca}. While this process explains certain young cold objects, it would overcool old ones unless an internal heating mechanism operates later. VCH naturally provides this late-time reheating pathway, maintaining elevated temperatures even in stars that once cooled rapidly.

In this paper, we present detailed neutron star cooling calculations incorporating both DUrca processes and VCH. Section~\ref{Sec:Methods} outlines the physical assumptions underlying our cooling calculations, including the basic principles of VCH and the caveats that arise when it operates alongside DUrca processes. We then describe our computational framework and physical inputs such as the adopted equations of state, superfluid gap models, and envelope physics. We also summarize the observational dataset used for comparison, consisting of isolated neutron stars with measured spin parameters. 
Section~\ref{Sec:Results} presents our results, combining validation against previous work \cite{Fujiwara_2024} with new calculations that test the validity of cooling models incorporating both VCH and DUrca processes. 
Based on these new calculations, we further identify conservative parameter ranges of $(B, P_0)$ for which the heating law $L_{\mathrm{h}} = J|\dot{\Omega}_\infty|$ remains applicable across different stellar masses. 
We then explore how the initial spin period $P_0$ and the surface magnetic flux density $B$ influence the thermal evolution of $1.4\,M_\odot$ and $2.0\,M_\odot$ stars.  In addition, we introduce a novel three-dimensional representation of cooling curves including magnetic-field effects, thereby mitigating potential ambiguities inherent in traditional two-dimensional plots.
Finally, Section~\ref{Sec:Conclusion} summarizes our main findings and outlines directions for future research, including the roles of magnetic-field evolution and rotational dynamics.

\section{Methods}\label{Sec:Methods}
\subsection{Vortex creep heating (VCH)}\label{Subsec:VCH}

\subsubsection{Basic principles of VCH}\label{Subsubsec:basic_principle_VCH}

\begin{figure}
    \centering
    \includegraphics[width=\linewidth]{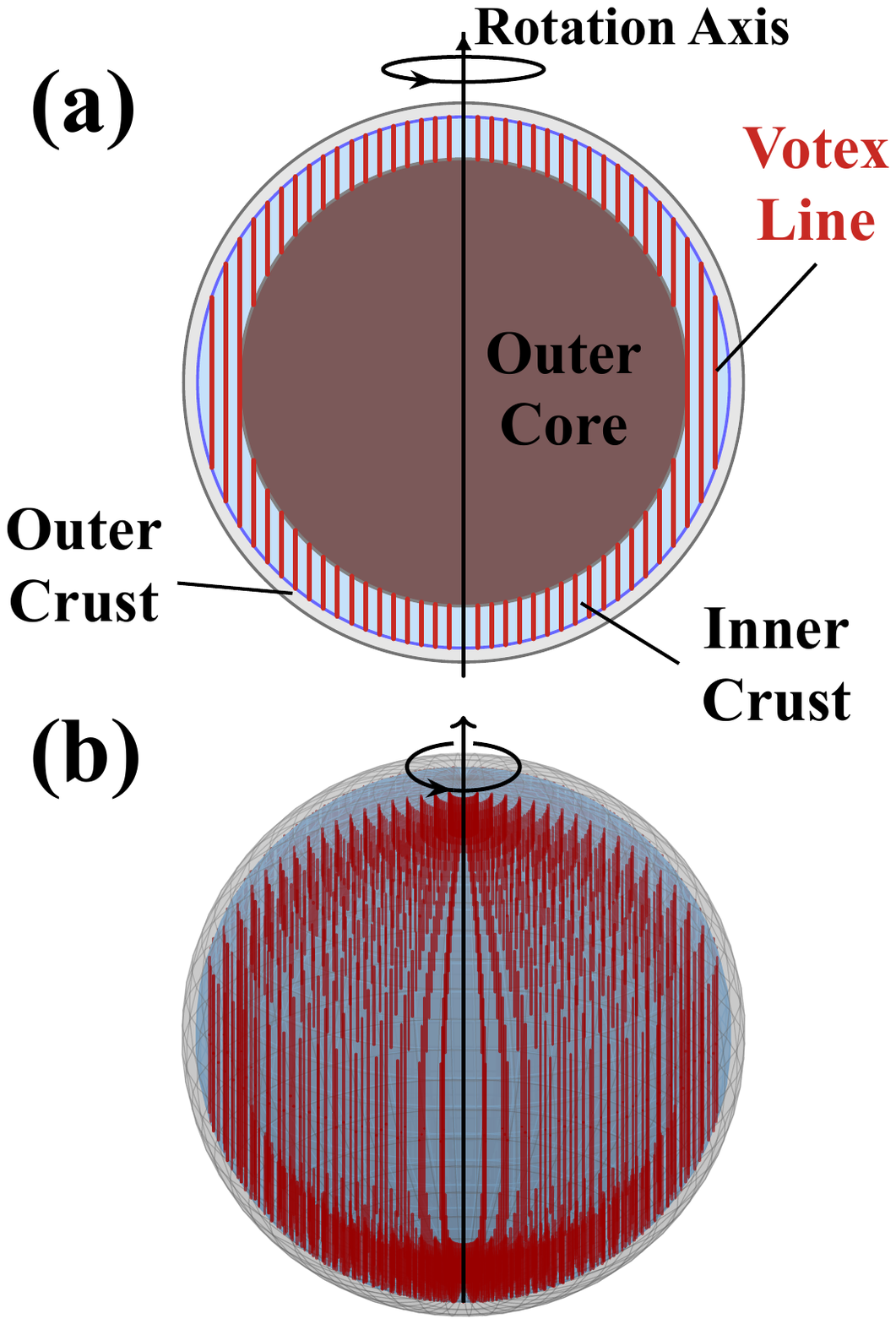}
    \caption{Schematic illustration of the two-component model.  
(a) Meridional cross section and (b) three-dimensional view.  
The \emph{crust} component consists of the solid outer crust, and the outer core with neutron ${}^3\mathrm{P}_2$ 
superfluid, which is expected to coexist with the 
proton superconductor and remain tightly coupled to the crustal 
motion 
\cite{Sauls_1982_core_is_crust_component_1,Alpar_1984_core_is_crust_component_2}.  
Quantized vortex lines (red) are aligned with the rotation axis 
and extend throughout the pinned superfluid region. Their outward 
creep in response to the external spin-down torque dissipates 
rotational energy, providing the physical basis for vortex creep 
heating.}
    \label{two-component-model}
\end{figure}

\begin{figure}
    \centering
    \includegraphics[width=0.8\linewidth]{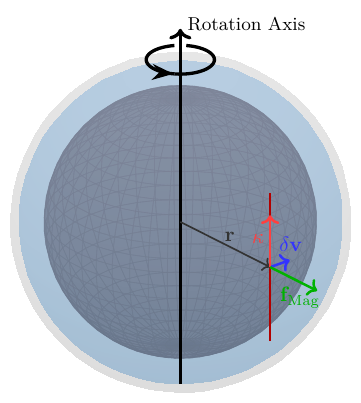}
    \caption{
Schematic depiction of the forces acting on a pinned vortex line in the inner crust.
A lag $\delta\Omega = \Omega_{\mathrm{s}} - \Omega_{\mathrm{c}}$ between the superfluid 
and the crustal rotation generates a relative azimuthal flow $\delta\mathbf{v}$ (blue), 
which produces an outward Magnus force $\mathbf{f}_{\rm Mag}$ (green) acting 
perpendicular to both $\delta\mathbf{v}$ and the vorticity vector $\boldsymbol{\kappa}$ (red).
The balance between $\mathbf{f}_{\rm Mag}$ and the pinning force $f_{\rm pin}$ defines
the critical lag $\delta\Omega_{\rm cr}$ that triggers vortex unpinning.
}
    \label{forces-diagram}
\end{figure}

The idea that superfluid vortices in the inner crust can dissipate rotational energy and heat a neutron star originates from two-component models in which a rigidly rotating \emph{crust} (ions plus tightly coupled charged fluids) is coupled to a neutron superfluid through pinned vortex lines (Fig.~\ref{two-component-model}) \cite{Baym_1969_two_component_model,Alpar_1984_VCH_1, Shibazaki_1989_VCH_2, Link_1996_VCH_3}. Note that the core neutron ${}^3\mathrm{P}_2$ superfluid is treated as part of the crustal component, 
because the neutron superfluid in this region is expected to coexist with the proton superconductor 
and to be magnetized through mutual interaction between neutrons and protons, and tightly coupled to the crustal component due to electron scatterings \cite{Sauls_1982_core_is_crust_component_1,Alpar_1984_core_is_crust_component_2}. The external torque directly brakes the crust, while the superfluid responds only via the creep of quantized vortices that interact with the nuclear lattice. We denote the crustal and superfluid angular velocities by $\Omega_{\mathrm{c}}(t)$ and $\Omega_{\mathrm{s}}(t,r)$, respectively. Pinning allows a lag $\delta\Omega\equiv \Omega_{\mathrm{s}}-\Omega_{\mathrm{c}}$ to build up, and the resulting relative flow exerts the Magnus force per unit length,
\begin{align}
\mathbf{f}_{\rm Mag} = n_\text{s}\,(\delta\boldsymbol{\Omega}\times\mathbf{r})\times\boldsymbol{\kappa},
\end{align}
with $|\boldsymbol{\kappa}|=\kappa=h/(2m_\text{n})$ the quantum of circulation \cite{Alpar_1984_VCH_1,Fujiwara_2024}. The geometry and force balance are illustrated in Fig.~\ref{forces-diagram}.
Balance between the outward Magnus force and the pinning force per unit length $f_{\rm pin}$ defines a critical lag,
\begin{align}
\delta\Omega_{\rm cr}(r)\,\equiv\, |\delta\Omega|_{f_\mathrm{Mag}=f_{\mathrm{pin}}}\,\simeq\, \frac{f_{\rm pin}(r)}{n_\text{s}(r)\,\kappa\,r},
\end{align}
which sets the threshold for unpinning \cite{Alpar_1984_VCH_1,Fujiwara_2024}.

Creep proceeds either by thermally activated hopping or by quantum tunneling across the pinning barrier 
\cite{Baym_1992_thermal_or_quantum_1, Link_1993_thermal_or_quantum_2}. 
Modeling the local pinning potential as a tilted periodic well yields the small-oscillation frequency near the potential minimum \cite{Fujiwara_2024},
\begin{align}
\omega_0 &\simeq \frac{\pi \kappa \Lambda}{4 R_{\rm WS}^2} 
\simeq 1.2\times10^{20}\,\mathrm{s}^{-1}
\left( \frac{R_{\mathrm{WS}}}{50\,\mathrm{fm}} \right)^{-2}
\left( \frac{\Lambda}{2} \right),
\end{align}
where $\Lambda$ ($2\lesssim \Lambda \lesssim10$ in the inner crust \cite{Link_1991_vortex_tension, Fujiwara_2024}) 
characterizes the vortex tension. 
The corresponding transition rate is given by 
\cite{Baym_1992_thermal_or_quantum_1, Link_1993_thermal_or_quantum_2}
\begin{align}
\mathcal{R}_{\rm VC} &\simeq \frac{\omega_0}{2\pi}
\exp\!\left(-\frac{|E_{\rm pin}|}{k_\mathrm{B} T_{\rm eff}}\right),
\end{align}
where \footnote{The subscript ``Q'' stands for ``Quantum-creep.'' 
This quantity corresponds to the transition temperature $T_{\mathrm{q}}$ 
used in earlier VCH studies (\textit{e.g.}, Ref.~\cite{Fujiwara_2024}), and the change 
in notation is introduced only to avoid confusion with particle-species labels.}
\begin{align}
k_\mathrm{B} T_{\rm eff}
&\equiv \frac{\hbar\omega_0}{2}
\coth\!\Big(\frac{T_{\mathrm{Q}}}{T}\Big)
\sim
\begin{dcases}
    k_{\mathrm{B}}T & (T\gg T_{\mathrm{Q}}),\\[2mm]
    \frac{\hbar \omega_0}{2} & (T \ll T_{\mathrm{Q}}),
\end{dcases}\label{def_teff}\\[2mm] 
T_{\mathrm{Q}} &\equiv \frac{\hbar\omega_0}{2k_{\mathrm{B}}}
\simeq 3.8\times10^8\,\mathrm{K}
\left(\frac{\omega_0}{10^{20}\,\mathrm{s}^{-1}}\right).
\end{align}
Thus creep is thermal for $T\gg T_{\mathrm{Q}}$ and quantum for $T\ll T_{\mathrm{Q}}$; with typical inner-crust parameters one finds $T_{\mathrm{Q}}\sim {\rm few}\times 10^8~\mathrm{K}$, implying quantum creep for old, cool stars \cite{Fujiwara_2024}.

When the creep rate becomes high enough to track the global spin-down, the system enters a steady regime in which the superfluid decelerates at the same rate as the crust, $\partial\Omega_{\mathrm{s}}/\partial t=\dot\Omega_{\mathrm{c}}\equiv\dot\Omega_\infty$, and the lag saturates near its critical value, $\delta\Omega_\infty\simeq\delta\Omega_{\rm cr}$ \cite{Shibazaki_1989_VCH_2, Link_1996_VCH_3, Fujiwara_2024}. The resulting frictional dissipation takes a simple global form,
\begin{align}
L_{\mathrm{h}} = J\,|\dot\Omega_\infty|,\qquad
J \equiv \int_{\rm pin} \delta\Omega_\infty\, \dd I_{\mathrm{p}},
\end{align}
where $\dd I_{\mathrm{p}}$ is the differential moment of inertia of the pinned region. In this stationary ``creep equilibrium,'' the parameter $J$ becomes effectively constant in time, as long as the structure of the pinned layer and the pinning configuration remain fixed. Therefore, $L_{\mathrm h}=J|\dot\Omega_\infty|$ provides a practical and robust prescription for the heating rate once the system has reached the quantum-creep steady state \cite{Fujiwara_2024}.

\subsubsection{Caveats when VCH operates alongside the DUrca process}\label{Subsubsec:caveats}

\begin{figure}
    \centering
    \includegraphics[width=\linewidth]{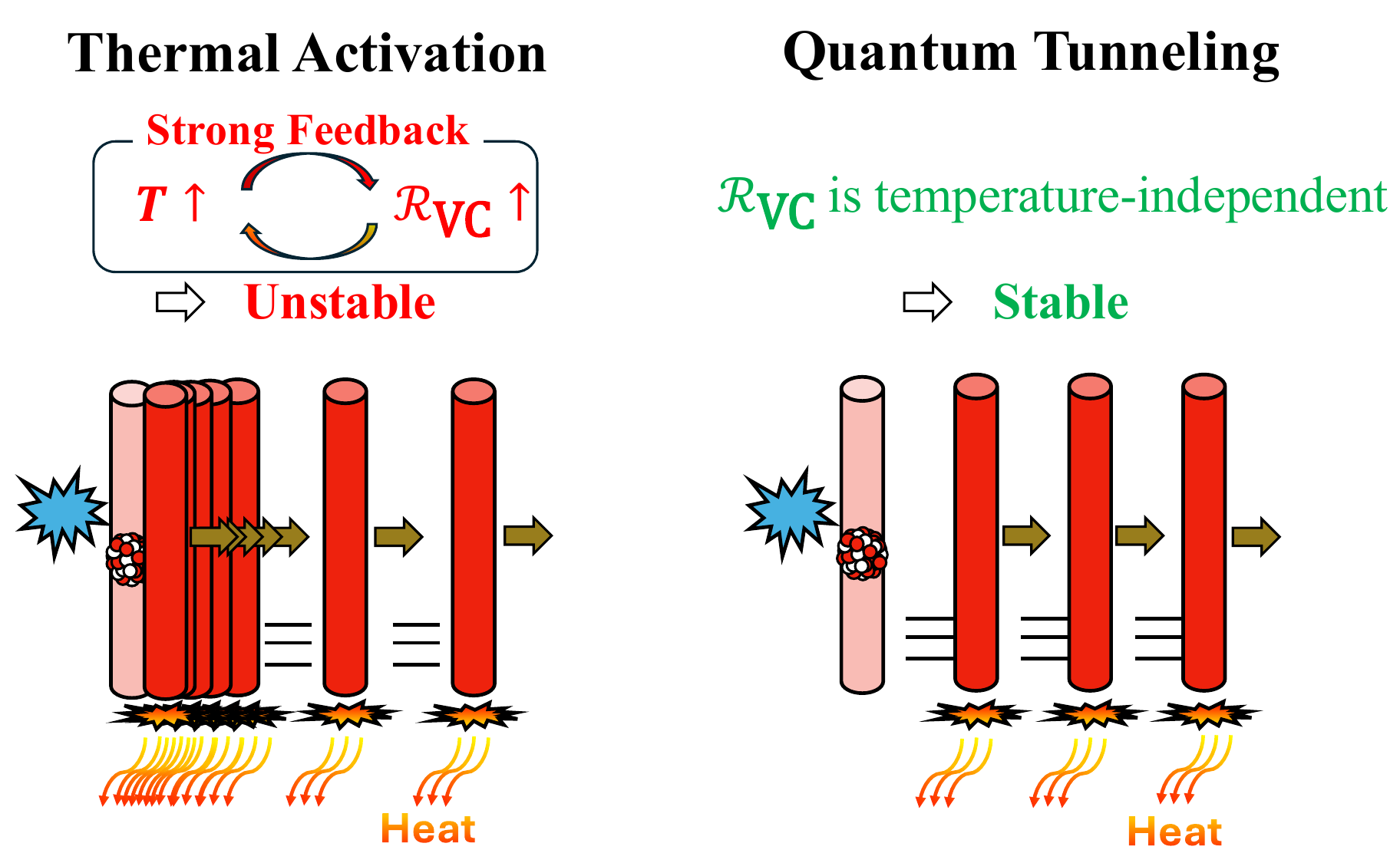}
    \caption{Comparison between thermally activated (left) and quantum-tunneling 
(right) vortex creep. In the thermally activated regime, the creep rate 
$\mathcal{R}_{\rm VC}$ grows exponentially with increasing temperature, enabling a 
positive feedback loop between heating and thermal evolution that may trigger 
thermo-rotational instabilities. In contrast, quantum tunneling yields a 
temperature-independent $\mathcal{R}_{\rm VC}$ and stabilizes the system once the 
superfluid has cooled sufficiently below $T_{\mathrm{Q}}$.
}
    \label{thermal_vs_quantum}
\end{figure}

In the evolutionary context, stars older than $\sim 10^5$~yr typically satisfy $T\ll T_{\mathrm{Q}}$ throughout most of the pinned inner crust, ensuring a stable quantum-creep equilibrium. By contrast, in massive stars where DUrca processes operate \cite{Lattimer_1991_durca}, rapid cooling via neutrino emissions at very early epochs ($t\sim 10^2$~yr) can produce strong temperature gradients \cite{Yakovlev_2004_neutron_star_cooling, Page_2006_the_cooling_of_compact_stars}, and thermal activation may remain locally relevant before the star fully enters the quantum regime. In this transitional regime, the creep velocity depends exponentially on temperature, $v_r\propto \exp[-E_{\rm pin}/(k_{\mathrm{B}}T)]$, which can establish a positive feedback between frictional heating and thermal evolution. Such coupling may cause thermo-rotational instabilities in which both temperature and spin-down rate oscillate quasi-periodically \cite{shibazaki_1995_instability_1, Larson_1998_instability_2}. These instabilities disappear once quantum tunneling dominates ($T\ll T_{\mathrm{Q}}$), rendering the system thermally stable at late times. The contrasting behavior of thermally activated and quantum-tunneling creep regimes is illustrated in Fig.~\ref{thermal_vs_quantum}.

A rigorous treatment of the thermally activated stage requires simultaneously solving the thermal evolution equations (discussed in Sec.~\ref{Sec:Cooling}) 
together with the coupled two-component dynamical equations \cite{Fujiwara_2024},
\begin{align}
I_{\mathrm{c}}\dot\Omega_{\mathrm{c}}(t) &= N_{\rm ext}(t) + N_{\rm int}(t), \\
\int \dd I_{\mathrm{p}}(r)\,\dot\Omega_{\mathrm{s}}(t,r) &= -N_{\rm int}(t),
\end{align}
where $N_{\rm int}$ is determined by the microphysics of vortex-nucleus interactions and by the temperature dependence of the pinning force $f_{\rm pin}$ \cite{Baym_1992_thermal_or_quantum_1}. 
Such \emph{thermo-rotational dynamics} models can capture transient feedback while being computationally demanding and dependent on uncertain microphysical parameters. 

In this work, we avoid such complexity and instead adopt a practical criterion 
for identifying the quantum-creep steady state, 
defined by the \textit{quantum-creep coverage fraction},
\begin{align}
f_\mathrm{Q}(t) &=
\frac{\displaystyle\int_{\rm pin}\Theta\!\big[\Delta_{\mathrm{Q}}(r,t)\le \epsilon\big]\, \dd I_\mathrm{p}}
{\displaystyle\int_{\rm pin}\Theta\!\big[T_\mathrm{c}^\mathrm{n}(r)-T(r,t)\big]\, \dd I_\mathrm{p}}, \label{def_f_Q}\\
\Delta_{\mathrm{Q}}(r,t)&\equiv\coth\!\Big(\frac{T_{\mathrm{Q}}(r)}{T(r,t)}\Big)-1,
\end{align}
where $T_\mathrm{c}^\mathrm{n}(r)$ is the critical temperature of neutron ${}^1\mathrm{S}_0$ superfluidity at radius $r$ inside the neutron star. Here $\dd I_\mathrm{p}$ denotes the local moment-of-inertia element within the pinned region, 
$\Theta$ is the Heaviside step function, 
and $\epsilon$ is a tolerance parameter that quantifies how closely the local state approaches 
the quantum-creep limit (\textit{i.e.}, how closely $T_{\mathrm{eff}}$ approaches $T_{\mathrm{Q}}$). If the denominator vanishes, which corresponds to no region satisfying $T<T_\mathrm{c}^\mathrm{n}$, we simply define $f_{\mathrm{Q}}=0$. We set $\epsilon = 0.01$ throughout this work, corresponding to a relative deviation 
$|T_{\mathrm{eff}}/T_{\mathrm{Q}} - 1| \lesssim 1\%$.

To clarify the definition of $\Delta_{\mathrm{Q}}$, recall that the effective local temperature governing the transition 
between thermally activated and quantum creep is given by Eq.~(\ref{def_teff}). 
By dividing Eq.~(\ref{def_teff}) by $k_{\mathrm{B}}T_{\mathrm{Q}} = \hbar\omega_0/2$, one obtains an equivalent form:
\begin{align}
T_{\rm eff} = T_{\mathrm{Q}}\,\coth\!\Big(\frac{T_{\mathrm{Q}}}{T}\Big).
\end{align}
The relative deviation of $T_{\rm eff}$ from $T_{\mathrm{Q}}$ is then
\begin{align}
\Delta_{\mathrm{Q}} \equiv \frac{T_{\rm eff}}{T_{\mathrm{Q}}}-1
              = \coth\!\Big(\frac{T_{\mathrm{Q}}}{T}\Big)-1.
\end{align}
Hence, $\Delta_{\mathrm{Q}}$ measures the local proximity to the quantum-creep regime: 
$\Delta_{\mathrm{Q}}\to0$ exponentially for $T\ll T_{\mathrm{Q}}$, 
while $\Delta_{\mathrm{Q}}\approx T/T_{\mathrm{Q}} - 1$ for $T\gg T_{\mathrm{Q}}$. 
Our criterion $\Delta_{\mathrm{Q}}\le\epsilon$ therefore selects the portions of the pinned region 
that have effectively entered the tunneling-dominated quantum-creep state. 
Throughout this work, the pinned region is treated as identical to the superfluid inner crust, 
where vortex pinning and creep are assumed to operate.

The resulting quantity $f_\mathrm{Q}$ represents the moment-of-inertia-weighted fraction of the pinned region 
that satisfies this quantum-creep condition, 
and its definition is physically consistent with that of the proportional constant $J$, 
which is likewise integrated over the inertial momenta in the effective pinning region. 
Because $\Delta_{\mathrm{Q}}\!\to\!0$ exponentially for $T\!\ll\!T_{\mathrm{Q}}$, 
$f_\mathrm{Q}\!\to\!1$ once most of the pinned region has reached the quantum-creep limit; 
beyond this point, the heating luminosity $L_{\mathrm{h}}=J|\dot{\Omega}_\infty|$ becomes effectively temperature-independent. 
Before this stage, especially in massive stars undergoing early DUrca cooling, 
$L_{\mathrm{h}}/L_\nu\ll1$, so any residual thermal dependence of the heating is negligible.

For completeness, we also account for the gradual activation of superfluidity itself 
through a secondary modulation factor, $\chi_{\rm sf}(t)$, 
which represents the moment-of-inertia-weighted fraction of the inner crust that has become superfluid:
\begin{align}
\chi_{\rm sf}(t)&=
\frac{\displaystyle\int_{\rm pin}\Theta\!\big[T_c^\text{n}(r)-T(r,t)\big]\, \dd I_\mathrm{p}}
{\displaystyle\int_{\rm pin} \dd I_\mathrm{p}},\label{def_chi_sf}
\end{align}
where $T_c^\text{n}(r)$ is the critical temperature of neutron ${}^1\mathrm{S}_0$ superfluidity at radius $r$ inside the neutron star. 
The total heating power is then expressed as
\begin{align}
L_{\mathrm{h}}(t)&=J\,|\dot\Omega_{\infty}(t)|\,\chi_{\rm sf}(t),\label{chi_sf_heating_power}
\end{align}
where $\chi_{\rm sf}(t)$ acts as a smooth switch that gradually turns on the heating 
as the ${}^1\mathrm{S}_0$ superfluid domain expands.
In practice, including $\chi_{\rm sf}$ mainly affects the early evolutionary phase 
when the crustal superfluid fraction is still small.
The quantity $f_\mathrm{Q}(t)$, on the other hand, is used as a diagnostic indicator to identify when the system has effectively entered the quantum-creep steady state, beyond which the heating luminosity $L_{\mathrm{h}} = J|\dot{\Omega}_\infty|$ becomes 
practically independent of temperature.

In the cooling simulations, the heating is implemented as a volumetric source within the inner crust, 
normalized so that its redshifted integral reproduces 
$L_{\mathrm{h}} = J|\dot\Omega_\infty|$ for a given $J$. 
Without explicitly introducing the detailed two-component dynamics or the microscopic pinning parameter $f_{\rm pin}$, 
we explore the macroscopic behavior of the VCH mechanism 
across a wide range of stellar conditions---including the stellar mass $M$, 
the envelope composition $\Delta M$, the surface magnetic flux density $B$, 
and the initial spin period $P_0$---while adopting the range of the proportional constant $J$ proposed in \cite{Fujiwara_2024}:
\begin{align}
J \simeq 10^{42.9\text{--}43.8}~\mathrm{erg\,s}.
\end{align}

\subsection{Neutron star model}\label{Subsec:Neutron-star-model}

The internal structure of a non-rotating, spherically symmetric neutron star is determined by the Tolman-Oppenheimer-Volkoff (TOV) equations \cite{Oppenheimer_1939_TOV,Shapiro_1983_BH_WD_NS_physics}, which describe the equilibrium between gravity and the pressure of dense nuclear matter. 
Assuming spherical symmetry, the radial profiles of pressure $P(r)$, enclosed mass $m(r)$, metric potential $\Phi(r)$, and baryon number density $n_\mathrm{B}(r)$ are obtained by integrating the following system:
\begin{align}
\frac{\dd m}{\dd r} &= 4\pi r^2\rho,\\
\frac{\dd P}{\dd r} &= -\frac{(\rho + P/c^2)\,(Gm + 4\pi G r^3 P/c^2)}{r^2(1 - 2Gm/c^2r)},\\
\frac{\dd\Phi}{\dd r} &= \frac{Gm/c^2 + 4\pi G r^3 P/c^4}{r^2(1 - 2Gm/c^2r)},\\
\frac{\dd a}{\dd r} &= \frac{4\pi r^2 n_\mathrm{B}}{\sqrt{1 - 2Gm/c^2r}},
\end{align}
where $G$ and $c$ are the gravitational constant and the speed of light, respectively. The boundary conditions require $m(0)=0$, and at the stellar surface $r=R$, the interior metric matches the exterior Schwarzschild solution,
\begin{align}
e^{\Phi(R)} = \sqrt{1 - \frac{2GM}{c^2R}},
\end{align}
where $M \equiv m(R)$ is the gravitational mass.

The solution of the TOV system depends critically on the equation of state (EoS), which provides the relation between the pressure $P$ and density $\rho$ of dense matter. 
In this work, we employ two representative nuclear EoSs for constructing neutron-star models. 
The APR (A18+$\delta v$+UIX$^{*}$) EoS \cite{Akmal_1998_APR_EoS} is included primarily for comparison with the results of the previous study \cite{Fujiwara_2024} 
to validate the correctness of our implementation. 
The main calculations, however, are based on the BSk24 EoS \cite{Pearson_2018_BSk24_EoS}, 
which belongs to the Brussels-Montreal family of unified functionals and provides a consistent description of the outer crust, inner crust, and core. 
BSk24 achieves an excellent global fit to experimental nuclear masses and reproduces astrophysical constraints from $\sim2\,M_\odot$ pulsar observations 
(\textit{e.g.}, Refs.~\cite{Demorest_2010_1.97_solarmass,Antoniadis_2013_2.01_solarmass}).

For the APR model, we use the tabulated EoS distributed with \texttt{NSCool} \cite{Page_2016_NSCool}, 
where the crust is described by the composite HZD-NV dataset \cite{Haensel_1989_HZD,Negele_1973_NV} 
and smoothly joined to the APR core, ensuring a consistent core-crust transition. 
This setup is standard in neutron-star cooling calculations and facilitates direct comparison with previous results. For the BSk24 model, we employ a unified tabulated form combining the outer-crust data (from Table~4 of Ref.~\cite{Pearson_2018_BSk24_EoS}) 
with the inner-crust and core segments generated using the public \texttt{bskfit18.f} routine \cite{Pearson_2018_BSk24_EoS,BSk_subroutine_website}. 
The tables are interpolated in the $(P,\rho)$ plane and coupled to our TOV solver written in Fortran90.

For each EoS, the TOV equations are integrated outward from a specified central density until the pressure vanishes at $r=R$, 
yielding the mass-radius relation and internal structure profiles. 
The resulting quantities---such as the core radius, crust thickness, and density stratification---serve as fixed inputs to the cooling and heating calculations presented in the following sections.

\subsection{Neutron star cooling}\label{Sec:Cooling}

The thermal evolution of a non-rotating, spherically symmetric neutron star is governed by the coupled energy-balance and heat-transport equations:
\begin{align}
\frac{\dd\!\left(L e^{2 \Phi}\right)}{\dd r} &=-\frac{4 \pi r^2 e^{\Phi}}{\sqrt{1-2 G m / c^2 r}}\!\left(C_V \frac{\dd T}{\dd t}+e^{\Phi}\!\left(Q_\nu-Q_\mathrm{h}\right)\!\right), \label{thermal_evolution_eq_1}\\
\frac{\dd\!\left(T e^{\Phi}\right)}{\dd r}&=-\frac{L e^{\Phi}}{4 \pi r^2 \lambda\, \sqrt{1-2 G m / c^2 r}},\label{thermal_evolution_eq_2}
\end{align}
where $L$ and $T$ are the local luminosity and temperature, $\lambda$ is the thermal conductivity, $C_V$ the specific heat per unit volume, and $Q_\nu$ and $Q_\mathrm{h}$ the neutrino emissivity and local heating rate. 
The metric function $\Phi(r)$ follows from the TOV solution. 
We assume freely escaping neutrinos and include no external heating sources 
other than the VCH described in Section~\ref{Subsec:VCH}.

Once the star reaches thermal relaxation, the nearly isothermal internal temperature distribution allows the local equations to be integrated into a global form \cite{Buschmann_2022}:
\begin{align}
    L_{\gamma}^{\infty} = -\,C\,\frac{\dd T_{\mathrm{b}}^{\infty}}{\dd t} - L_{\nu}^{\infty} + L_{\mathrm{h}}^{\infty},
\end{align}
where $C$ is the total heat capacity, and $L_{\nu}^{\infty}$ and $L_{\mathrm{h}}^{\infty}$ are the redshifted neutrino and internal-heating luminosities, respectively. 
This formulation highlights the interplay between neutrino losses, photon emission, and internal heating.

Here, we adopt the \textit{barotropic EoS approximation} (see, \textit{e.g.}, Refs.~\cite{Gnedin_2001,Page_2016_NSCool}), which exploits the strong degeneracy of dense matter to decouple the mechanical and thermal structures. The cooling region is defined for densities above a fiducial boundary $\rho_\mathrm{b} = 10^{10}\,\mathrm{g\,cm^{-3}}$, 
where the matter is sufficiently degenerate. 
The stellar metric and mass profiles $(m(r),\Phi(r))$ are computed from the TOV equations for the chosen EoS. 
Below this boundary, the heat flow through the outer envelope determines the observable surface emission.

The envelope acts as a thermal regulator that connects the internal temperature at $\rho_\mathrm{b}$ to the surface value $T_\mathrm{s}$. 
We employ the $T_\mathrm{s}$-$T_\mathrm{b}$ relation in Ref.~\cite{Potekhin_1997_accreted_envelope} 
and represent the composition using the light-element mass parameter $\Delta M$. 
Larger $\Delta M$ corresponds to a thicker carbon layer and thus a higher surface temperature for the same interior $T_\text{b}$. We explore the range 
\[
10^{-18}\,M_\odot \le \Delta M \le 10^{-7}\,M_\odot.
\]

The photon luminosity observed at infinity (\textit{i.e.}, by a distant observer) is
\begin{align}
  L_{\gamma}^{\infty} = 4\pi R^2 \sigma_{\mathrm{SB}} T_{\mathrm{eff}}^4
  \!\left(1 - \frac{2GM}{c^2R}\right),
\end{align}
where $T_{\mathrm{eff}}^{\infty} = T_{\mathrm{eff}}\sqrt{1 - 2GM/c^2R}$ and 
$R_{\infty} = R/\sqrt{1 - 2GM/c^2R}$ \cite{Thorne_1977}. 
These relations provide the direct connection between theoretical cooling curves and observable quantities.

The time evolution is computed with our one-dimensional, general-relativistic cooling code 
developed in Fortran90 developed recently \cite{Nam_2025_data_driven_cas_a_ns}, which follows the numerical framework of \texttt{NSCool} \cite{Page_2016_NSCool}. The code employs an implicit finite-difference scheme to advance Eqs.~\eqref{thermal_evolution_eq_1} and \eqref{thermal_evolution_eq_2} from an initially 
isothermal configuration ($T e^{\Phi}=10^{10}\,\mathrm{K}$). The adopted microphysics follows 
standard prescriptions (see, \textit{e.g.}, Ref.~\cite{Potekhin_2015_review_physical_input}), using the 
SFB neutron $^1\mathrm{S}_0$ gap \cite{Schwenk_2003_SFB}, the CCDK proton $^1\mathrm{S}_0$ gap 
\cite{Elgaroy_1996_ccdk}, and the TToa neutron $^3\mathrm{P}_2$ gap \cite{Takatsuka_2004_nt_TTav_TToa}.

For the neutron ${}^3\mathrm{P}_2$ PBF process, we adopt the 
standard axial-vector reduction factor $q = 0.76$, rather than the anomalous contribution 
$q \simeq 0.19$ reported in Ref.~\cite{Leinson_2010}. The applicability of the anomalous term may 
depend on uncertain in-medium effects, and we retain the conventional treatment to ensure 
consistency with Ref.~\cite{Fujiwara_2024} and other comparative works.

VCH is incorporated as an internal energy source confined to the inner crust, 
implemented consistently with the prescription described in Sec.~\ref{Subsec:VCH}. 
For further details of our numerical implementation, see Ref.~\cite{Nam_2025_data_driven_cas_a_ns}.

\subsection{Superfluid and superconducting gap models}\label{sf_gap_model}

The critical temperature for baryon superfluidity or superconductivity is related to the pairing gap $\Delta$ through the standard BCS expression,
\begin{align}
k_\mathrm{B}T_{\mathrm{c}}\approx
\begin{dcases}
        0.5669\,\Delta, & \text{singlet (isotropic) pairing,}\\[2mm]
        0.5669\,\dfrac{\Delta}{\sqrt{8\pi}}, & \text{triplet (anisotropic) pairing,}
\end{dcases}
\label{pairing_gap_temp_relation}
\end{align}
where $k_\mathrm{B}$ is the Boltzmann constant.  
This convention is widely adopted in neutron-star cooling studies (see, \textit{e.g.}, Ref.~\cite{Ho_2015}).

The following parametrization has been widely used to represent the density dependence of the pairing gap \cite{Kaminker_2001_parametrization}:
\begin{align}
    \Delta(\kfx, T=0) 
    = \Delta_0  
      \frac{(\kfx - k_0)^2}{(\kfx - k_0)^2+k_1}
      \frac{(\kfx - k_2)^2}{(\kfx - k_2)^2+k_3},
    \label{traditional_parametrization}
\end{align}
where $\mathrm{x}\in\{n,p\}$ denotes the baryon species, and $\Delta_0$, $k_0$, $k_1$, $k_2$, and $k_3$ are fitting parameters.  
The corresponding parameter sets for various singlet and triplet pairing models are summarized in Table~II of Ref.~\cite{Ho_2015}.

Recently, we have proposed a new parametrization with a more transparent physical interpretation in our previous work \cite{Nam_2025_data_driven_cas_a_ns}. Although functionally equivalent to the conventional form in Eq.~\eqref{traditional_parametrization}, the new representation explicitly relates the peak height, position, width, and asymmetry of the gap, facilitating direct comparison between models and enabling data-driven optimization of the $^3\mathrm{P}_2$ neutron gap from observations (see Ref.~\cite{Nam_2025_data_driven_cas_a_ns} for details). For consistency with the present analysis, we adopt this new parametrization throughout this study.

\subsection{Observational data of isolated cooling neutron stars}\label{obda}

To validate our theoretical cooling calculations, we compare the results with observational data of neutron-star surface temperatures compiled in the publicly available Ioffe Institute database \cite{Potekhin_data_of_ioslated_ns}. 
This database, maintained by Potekhin and collaborators, provides a comprehensive and regularly updated catalog of thermal-emission properties for isolated neutron stars, including the spin period $P$, its derivative $\dot{P}=dP/dt$, the dipolar surface magnetic flux density $B_{\mathrm{dip}}=3.2\times10^{19}\sqrt{P\dot P}$~G derived under the canonical vacuum-dipole model \cite{Ostriker_1969_dipole_radiation}, the characteristic age $t_\mathrm{c}=P/(2\dot P)$, the independently inferred kinematic age $t_*$, and the redshifted effective temperature $T_{\infty}$ (in eV) as measured by a distant observer. 
For direct comparison with theoretical models, we adopt $T_{\infty}$ values obtained from thermal X-ray spectra with well-constrained blackbody or atmosphere fits. 
The specific sources used in this study are summarized in Table~\ref{observational_data_updated}.

Following the classification scheme in Ref.~\cite{Potekhin_2020_data_of_isolated_ns}, isolated neutron stars can be broadly categorized according to their surface magnetic flux densities and emission properties. 
\textit{Weakly magnetized thermal emitters}, including Central Compact Objects (CCOs) and thermally emitting isolated neutron stars (TINSs), exhibit soft X-ray radiation and relatively weak magnetic fields ($B_{\mathrm{dip}}<5\times10^{11}\,\mathrm{G}$). 
\textit{Ordinary pulsars} possess moderate fields ($B_{\mathrm{dip}}\sim10^{12\text{--}13}\,\mathrm{G}$) and typically show well-defined thermal X-ray components, exemplified by the “Three Musketeers.” 
\textit{High-B pulsars} ($B_{\mathrm{dip}}\sim10^{13\text{--}14}\,\mathrm{G}$) display non-uniform surface temperatures but comparable bolometric luminosities. 
The \textit{Magnificent Seven}, discovered in the ROSAT All-Sky Survey (see, \textit{e.g.}, Refs.~\cite{Haberl_2007_rosat_review_1, Turolla_2009_rosat_reveiw_2, Kaplan_2009_rosat_review_3}, for reviews), exhibit nearly pure thermal spectra despite magnetic fields similar to high-B pulsars. 
The \textit{Upper-Limit} class consists of young pulsars with only upper bounds on thermal luminosity, while a subset of sources with small emitting areas ($R_{\mathrm{eff}}^{\infty}\!\lesssim\!0.5\,\mathrm{km}$) is interpreted as \textit{hot spots} associated with polar-cap heating~\cite{Potekhin_2020_data_of_isolated_ns}.

Although the database covers a wide variety of neutron-star populations, our theoretical framework necessarily involves several simplifying assumptions that constrain the scope of direct quantitative comparison:  
(i) magnetic-field effects on the envelope and crust are neglected;  
(ii) the barotropic EoS approximation limits the microphysical detail of the outer-crust and envelope treatment \cite{Potekhin_2018}; and  
(iii) spherical symmetry excludes rapid rotation and magnetically induced deformation.  
Accordingly, our quantitative analysis focuses on \textit{ordinary pulsars}, for which these approximations remain well justified and theoretical cooling curves can be meaningfully compared with observations. Nevertheless, for completeness, we also display other neutron-star classes in the figures to provide qualitative context and facilitate broader comparison across different populations.

\begin{table*}[htp]
\centering
\small
\caption{
Classification and observational parameters of isolated neutron stars adopted in this study. Data are compiled from the Ioffe Institute online catalog \cite{Potekhin_data_of_ioslated_ns}. A corresponding $P$--$\dot{P}$ diagram for the subset with measured $(P,\dot{P})$ is shown in Fig.~\ref{P_Pdot_diagram} (see Appendix~\ref{Appendix:PPdot}).}
\label{observational_data_updated}
\begin{tabular}{llcccccc}
\hline
No. & Name & $P$ & $\dot{P}$ & $B_{\rm dip}$ & $t_c$ & $t_*$ & $T_{\infty}$ \\
& & (s) & ($10^{-15}$ s/s) & ($10^{12}$ G) & (kyr) & (kyr) & (eV) \\
\hline
\multicolumn{8}{l}{\textbf{I. Weakly magnetized thermal emitters}} \\
1  & RX J0822.0$-$4300 (Puppis A)     & 0.113  & 0.00928 & 0.029  & 254000 & 3.7-5.2 & 240-475 \\
2  & CXOU J085201.4$-$461753 (Vela Jr.) & --     & --      & --     & --                  & 2.4-5.1 & 90$\pm$10 \\
3  & 2XMM J104608.7$-$594306           & --     & --      & --     & --                  & 11-30   & 40-70 \\
4  & 1E 1207.4$-$5209                  & 0.424  & 0.0223    & 0.098  & 301000 & 2.333-21.0 & 90-250 \\
5  & 1RXS J141256.0+792204 (“Calvera”) & 0.0592 & 3.29    & 0.45   & 285                 & --       & 101-110 \\
6  & CXOU J160103.1$-$513353           & --     & --      & --     & --                  & 0.6-1.0 & 125-147 \\
7  & 1WGA J1713.4$-$3949               & --     & --      & --     & --                  & 1.608-1.609    & 128-145 \\
8  & XMMU J172054.5$-$372652           & --     & --      & --     & --                  & 0.6-0.7 & $162\pm6$ \\
9  & XMMU J173203.3$-$344518           & --     & --      & --     & --                  & 2-6      & $156\pm6$ \\
10 & CXOU J181852.0$-$150213           & --     & --      & --     & --                  & 2.7-6.0  & $130\pm20$ \\
11 & CXOU J185238.6+004020 (Kes 79)    & 0.105  & 0.00868 & 0.031  & 192000 & 3.2-7.8  & $133\pm 1$ \\
12 & CXOU J232327.8+584842 (Cas A)     & --     & --      & --     & --                  & 0.3195-0.3387 & 120-128 \\
\hline
\multicolumn{8}{l}{\textbf{II. Ordinary pulsars}} \\
13 & PSR J0205+6449 (3C 58)            & 0.0657 & 194     & 3.6    & 5.37      & 0.8194-0.8196 & 43-54 \\
14 & PSR J0357+3205 (“Morla”)          & 0.444  & 13.0    & 2.4    & 541       & 200-1300 & 30-45 \\
15 & PSR J0538+2817 (S147)             & 0.143  & 3.67    & 0.73   & 620       & 20-60   & $91\pm5$ \\
16 & PSR J0554+3107                    & 0.465  & 143     & 8.2    & 51.7      & --      & $48\pm3$ \\
17 & PSR J0633+0632                    & 0.297  & 79.6    & 4.9    & 59.2      & --      & $53\pm4$ \\
18 & PSR J0633+1746 (“Geminga”)        & 0.237  & 1.10    & 1.6    & 342       & --      & $42\pm2$ \\
19 & PSR B0656+14 (Monogem)            & 0.385  & 54.9    & 4.7    & 111       & --      & $68\pm1$ \\
20 & PSR B0833$-$45 (Vela)             & 0.0893 & 125     & 3.4    & 11.3      & 17-27   & 56-60 \\
21 & PSR B0950+08                      & 0.253  & 0.230   & 0.24   & 17500 & 600-7600 & 5-11 \\
22 & PSR B1055$-$52                    & 0.197  & 5.83    & 1.1    & 535       & --      & $68\pm3$ \\
23 & PSR J1357$-$6429                   & 0.166  & 360     & 7.8    & 7.31      & --      & $64\pm4$ \\
24 & PSR B1706$-$44                    & 0.102  & 93.0    & 3.1    & 17.5      & --      & 40-230 \\
25 & PSR J1740+1000                    & 0.154  & 21.5    & 1.8    & 114       & --      & 70-140 \\
26 & PSR J1741$-$2054                   & 0.414  & 17.0    & 2.7    & 386       & --      & $60\pm2$ \\
27 & PSR B1822$-$09                    & 0.769  & 52.5    & 6.4    & 233       & --      & $83\pm4$ \\
28 & PSR B1823$-$13                    & 0.101  & 75.3    & 2.8    & 21.4      & --      & 92-101 \\
29 & PSR J1836+5925 (“Next Geminga”)   & 0.173  & 1.50    & 0.52   & 1830 & -- & 13.7-19.2 \\
30 & PSR B1951+32 (CTB 80)             & 0.0395 & 5.85    & 0.49   & 107       & 46-82   & $130\pm20$ \\
31 & PSR J1957+5033                    & 0.375  & 7.00    & 1.6    & 840       & --      & 20.1-25.3 \\
32 & PSR J2021+3651                    & 0.104  & 95.7    & 3.2    & 17.2      & --      & 58-69 \\
33 & PSR B2334+61 (G114.3+00.3)        & 0.495  & 193     & 9.9    & 40.6      & 7.6-7.8 & 30-52 \\
\hline
\multicolumn{8}{l}{\textbf{III. High-B pulsars}} \\
34 & PSR J0726$-$2612                   & 3.442  & 293     & 32     & 186       & --      & 63-80 \\
35 & PSR J0837$-$2454                   & 0.629  & 349     & 15     & 28.6      & --      & 27-70 \\
36 & PSR J1119$-$6127                   & 0.408  & 4020    & 41     & 1.60      & 4.2-7.1 & 80-210 \\
37 & PSR B1509$-$58                     & 0.151  & 1530    & 15     & 1.56      & --      & 133-149 \\
38 & PSR J1718$-$3718                   & 3.379  & 1610    & 75     & 33.2      & --      & 57-200 \\
39 & PSR J1819$-$1458                   & 4.263  & 563     & 50     & 120       & --      & 113-140 \\
\hline
\multicolumn{8}{l}{\textbf{IV. The Magnificent Seven}} \\
40 & RX J0420.0$-$5022                  & 3.453  & 29.1    & 10     & 1880      & --      & 45.0$\pm$2.6 \\
41 & RX J0720.4$-$3125                  & 8.391  & 71.5    & 25     & 1860      & 700-1000 & 91.4-93.3 \\
42 & RX J0806.4$-$4123                  & 11.370 & 10.6    & 11     & 17000 & -- & 90-110 \\
43 & RX J1308.6+2127                    & 10.312 & 112     & 34     & 1460      & 300-800 & 50-90 \\
44 & RX J1605.3+3249                    & --     & --      & --     & --        & 380-510 & 35-120 \\
45 & RX J1856.5$-$3754                  & 7.055  & 30.1    & 15     & 3720      & 340-500 & 36-63 \\
46 & RX J2143.0+0654                    & 9.428  & 41.0    & 20     & 3700      & --      & 40-104 \\
\hline
\multicolumn{8}{l}{\textbf{V. Upper limits}} \\
47 & PSR J0007+7303 (CTA 1)             & 0.316  & 360     & 1.1    & 13.9      & 9.1-9.3 & $<$200 \\
48 & PSR B0531+21 (Crab)                & 0.0334 & 421     & 3.8    & 1.26      & 0.9535-0.9537 & $<$180 \\
49 & PSR B1727$-$47 (RCW 114)           & 0.830  & 164     & 12     & 80.5      & 40-60  & $<$33 \\
50 & PSR J2043+2740                     & 0.0961 & 1.27    & 0.35   & 1200      & --      & $<$80 \\
51 & PSR B2224+65 (Guitar)              & 0.683  & 9.66    & 2.6    & 1130      & --      & $<$112 \\
\hline
\multicolumn{8}{l}{\textbf{VI. Hot spots}} \\
52 & PSR B0114+58                       & 0.101  & 5.85    & 0.78   & 275       & --      & $170\pm20$ \\
53 & PSR B0943+10                       & 1.098  & 3.49    & 2.0    & 4980      & --      & 79-220 \\
54 & PSR B1133+16                       & 1.188  & 3.73    & 2.1    & 5040      & --      & 160-230 \\
55 & PSR J1154$-$6250                   & 0.282  & 0.559   & 0.40   & 7990      & --      & $210\pm40$ \\
56 & PSR B1929+10                       & 0.227  & 1.16    & 0.52   & 3110      & --      & $300\pm30$ \\
\hline
\end{tabular}
\end{table*}

\section{Results and Discussion}\label{Sec:Results}

\subsection{Verification of the implementation of the VCH}\label{Sec:verification}

\subsubsection{Validation against the previous study}

\begin{figure}[t]
    \centering
    \includegraphics[width=1.0\linewidth]{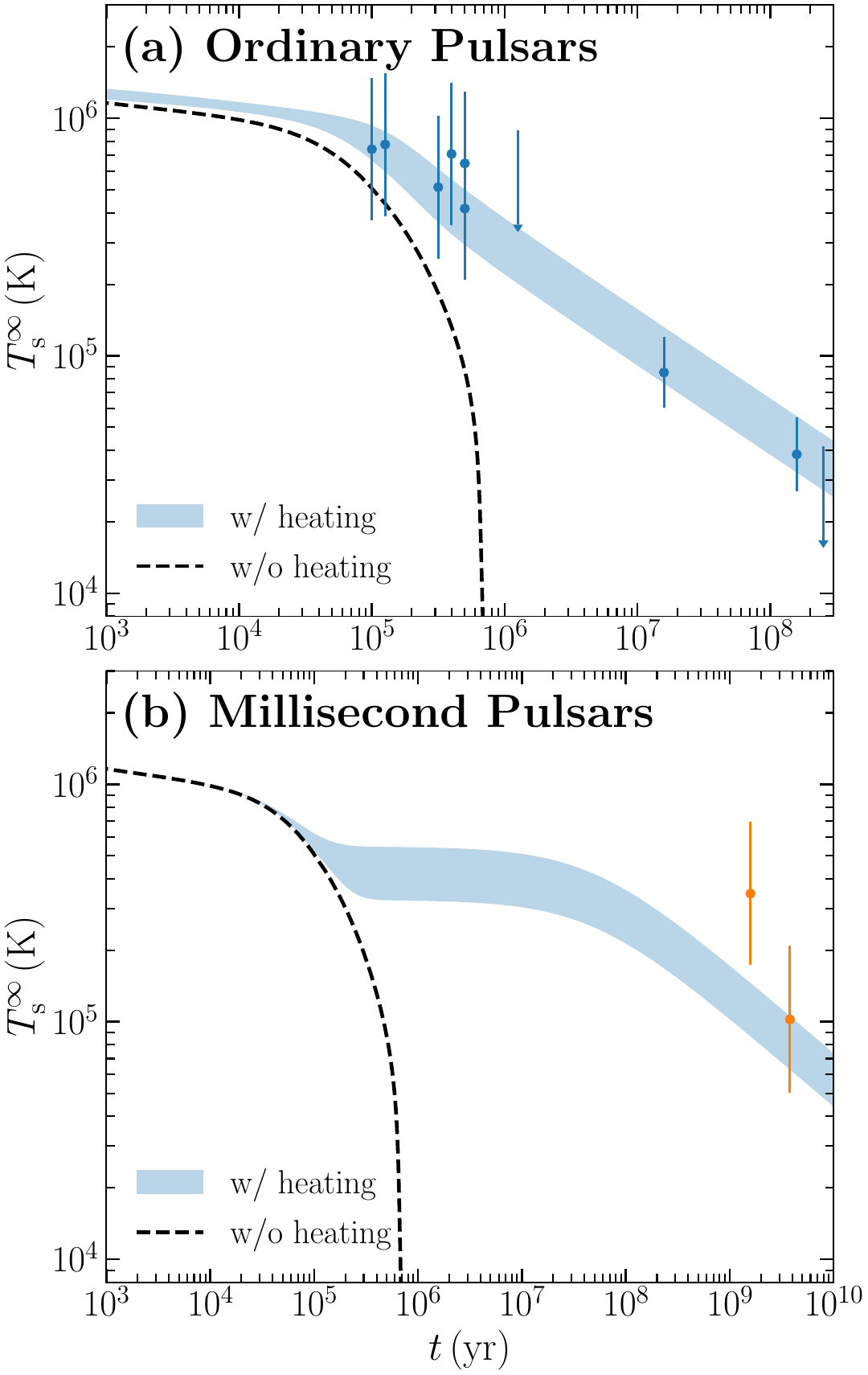}
    \caption{(a) Ordinary pulsars and (b) millisecond pulsars. Surface temperature evolution with (blue bands) and without (black dashed) vortex creep heating (VCH). The blue bands correspond to $J \simeq 10^{42.9\text{--}43.8}\,$erg\,s. Observational data from Fig.~5 of Ref.~\cite{Fujiwara_2024} are shown with the same color scheme. Our results successfully reproduce the VCH cooling curves presented in Fig.~6 of Ref.~\cite{Fujiwara_2024}.}
    \label{ordinary_vs_millisecond}
\end{figure}

Building on the previous investigation of VCH in neutron stars by Fujiwara \textit{et al}.~\cite{Fujiwara_2024}, we implemented the same heating mechanism within our computational framework. As an initial validation, we reproduced the results reported in their study, using the APR \cite{Akmal_1998_APR_EoS} core EoS with the HZD-NV \cite{Haensel_1989_HZD, Negele_1973_NV} crust EoS as described in Sec.~\ref{Subsec:Neutron-star-model}, ensuring a fully consistent comparison with Ref.~\cite{Fujiwara_2024}. Figure~\ref{ordinary_vs_millisecond} demonstrates a high level of consistency with Fig.~6 in Ref.~\cite{Fujiwara_2024}, confirming the successful incorporation of the VCH model into our code.

The blue bands in Fig.~\ref{ordinary_vs_millisecond} show the range of cooling curves obtained by varying the proportional constant $J$ within $J \simeq 10^{42.9\text{--}43.8}\,$erg\,s. Here, $J$ quantifies the coupling strength between the superfluid neutrons and the normal component. The lower and upper edges of the bands correspond to the minimum and maximum values of $J$, respectively, and cooling curves for any $J$ within this interval lie inside the shaded region. For comparison, cooling without VCH is shown by black dashed lines.

In addition to the variation in $J$, the cooling behavior depends on the rotational evolution of the star. For both ordinary and millisecond pulsars, we adopt the same spin parameters as those used in Ref.~\cite{Fujiwara_2024}. For ordinary pulsars [Fig.~\ref{ordinary_vs_millisecond}(a)], we take $P\dot{P} = 10^{-15}\,\mathrm{s}$ and an initial spin period of $P_0 = 10\,\mathrm{ms}$, assuming that magnetic dipole radiation dominates the external torque. For millisecond pulsars [Fig.~\ref{ordinary_vs_millisecond}(b)], we follow the representative case of PSR~J2124--3358, adopting $P\dot{P} = 3.3\times10^{-22}\,\mathrm{s}$ and $P_0 = 1\,\mathrm{ms}$ \cite{Fujiwara_2024}.

In both cases, the cooling curves including VCH deviate markedly from standard predictions without heating at approximately $t \sim 10^5$ years. The sustained heating maintains higher surface temperatures at later times, leading to significantly better agreement with observational data---particularly for older neutron stars, for which conventional cooling models tend to underpredict the thermal luminosity. These results clearly demonstrate that VCH plays an essential role in accurately modeling the long-term thermal evolution of neutron stars.

Although Ref.~\cite{Fujiwara_2024} explored both ordinary and millisecond pulsars, having established the consistency of our implementation with their results, we now focus exclusively on ordinary pulsars in the subsequent sections. This restriction is motivated by the fact that our cooling code assumes spherical symmetry and models isolated neutron stars, whereas millisecond pulsars are rapidly rotating systems that are typically spun up through accretion in binary environments, potentially violating this assumption.

\subsubsection{Constraints ensuring computational validity}

Having validated our implementation against Ref.~\cite{Fujiwara_2024} using the APR-based setup, we now focus on the main results of this work. Unless otherwise specified, all subsequent thermal evolution and VCH 
calculations are performed using the unified BSk24 \cite{Pearson_2018_BSk24_EoS} EoS, which provides a consistent 
core-crust description and satisfies current astrophysical mass constraints.

\begin{figure}[t]
\centering
\includegraphics[width=1.0\linewidth]{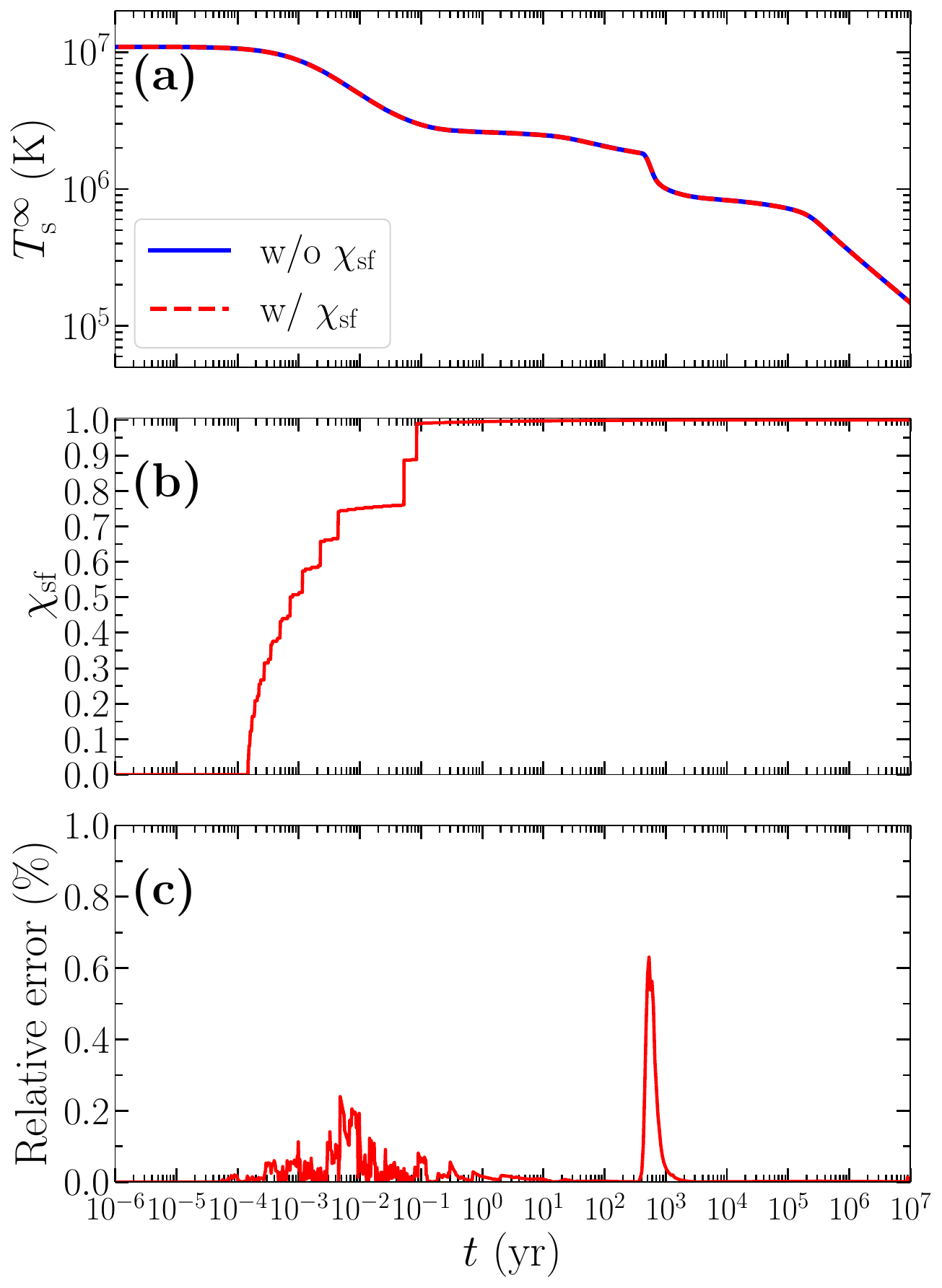}
\caption{Effect of including the fraction $\chi_{\mathrm{sf}}$ in the heating luminosity.
(a) Red dashed (with $\chi_{\mathrm{sf}}$) and blue solid (without $\chi_{\mathrm{sf}}$) curves show the surface temperature evolution $T_{\mathrm{s}}^\infty(t)$;
(b) Time evolution of $\chi_{\mathrm{sf}}$, representing the fraction of the moment of inertia occupied by superfluid neutrons in the inner crust;
(c) Relative temperature difference ${|T_{\mathrm{w/}\,\chi_{\mathrm{sf}}}-T_{\mathrm{w/o}\,\chi_{\mathrm{sf}}}|}/{T_{\mathrm{w/o}\,\chi_{\mathrm{sf}}}}$ between the two cases.
The maximum discrepancy is only $\sim0.65\%$, confirming that $\chi_{\mathrm{sf}}\rightarrow1$ early enough to justify the simplified implementation $L_{\mathrm{h}}=J|\dot{\Omega}_\infty|$ in practical applications.}
\label{chi_effect}
\end{figure}

As a first step, we examine the impact of applying the fraction $\chi_{\mathrm{sf}}$ defined in Eq.~\eqref{def_chi_sf}. Recall that $\chi_{\mathrm{sf}}$ represents the fraction of the moment of inertia weighted by the region where neutron $^1\mathrm{S}_0$ superfluidity forms in the inner crust, such that the heating luminosity becomes $L_{\mathrm{h}} = J|\dot{\Omega}_\infty(t)|\chi_{\mathrm{sf}}(t)$ as in Eq.~\eqref{chi_sf_heating_power}. Figure~\ref{chi_effect} compares cooling curves with and without multiplying $\chi_{\mathrm{sf}}$ for $P\dot{P} = 10^{-15}\,\mathrm{s}$, $P_0 = 30$~ms, $J=10^{43.8}~\mathrm{erg\,s}$, $M = 1.4\,M_\odot$, and an iron envelope ($\Delta M=0$). The maximum relative temperature deviation is only $\sim$0.65\%, indicating that the simplification $\chi_{\mathrm{sf}} \to 1$ is well justified in practical applications.

\begin{figure}[t]
\centering
\includegraphics[width=1.0\linewidth]{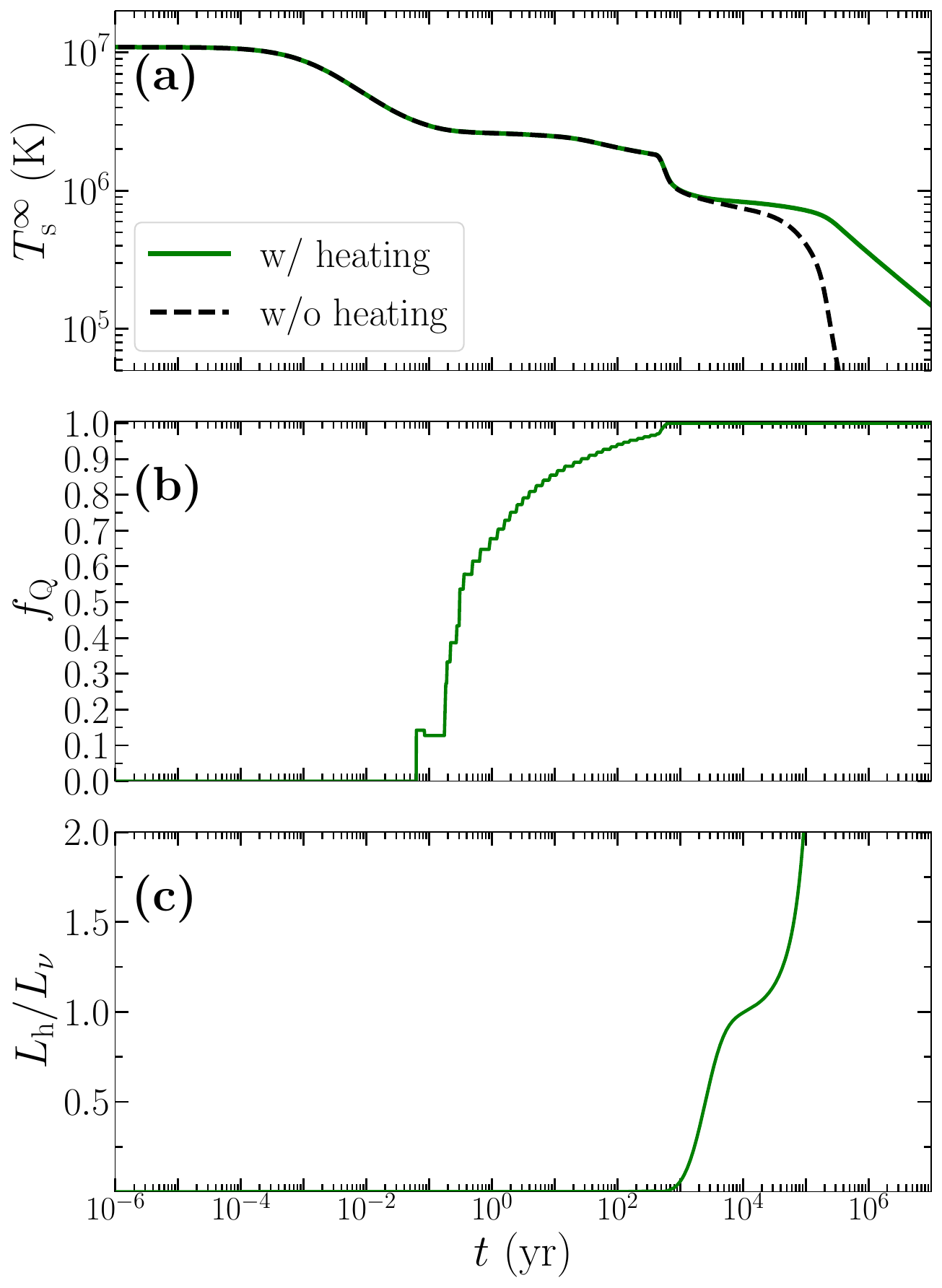}
\caption{Evolution of the quantum-creep coverage fraction $f_{\mathrm{Q}}$ and its relation to vortex creep heating (VCH).
(a) Cooling curves with (green solid) and without (black dashed) VCH;
(b) Growth of $f_{\mathrm{Q}}$, where $f_{\mathrm{Q}}\rightarrow1$ signifies that quantum tunneling dominates vortex creep;
(c) Ratio of VCH luminosity to neutrino luminosity, $L_{\mathrm{h}}/L_\nu$, illustrating when heating begins to significantly influence cooling curves.
In this case, heating becomes effective only after entering the quantum-creep regime ($f_{\mathrm{Q}}\approx1$).}
\label{quantum_creep_region_1.4M}
\end{figure}

Next, we analyze the time evolution of the quantum-creep coverage fraction $f_{\mathrm{Q}}$ defined in Eq.~\eqref{def_f_Q}. Figure~\ref{quantum_creep_region_1.4M} shows that $f_{\mathrm{Q}}$ reaches unity only after $\sim10^3$~yr, when vortex creep is governed by quantum tunneling. This marks the point at which the steady-state assumption becomes valid and VCH begins to significantly alter cooling curves.

The onset timing of VCH influence depends on the stellar mass and spin parameters. In some cases, heating may affect cooling before $f_{\mathrm{Q}}\rightarrow1$, violating the steady-state assumption. To quantify this, we define $t_{\mathrm{sep}}$ as the time when the relative deviation between heating and no-heating cooling curves first exceeds 1\%, and $t_{\mathrm{Q}}$ as the time when $f_{\mathrm{Q}}>0.99$. The condition $t_{\mathrm{sep}} > t_{\mathrm{Q}}$ ensures calculational validity.

\begin{figure}[t]
\centering
\includegraphics[width=1.0\linewidth]{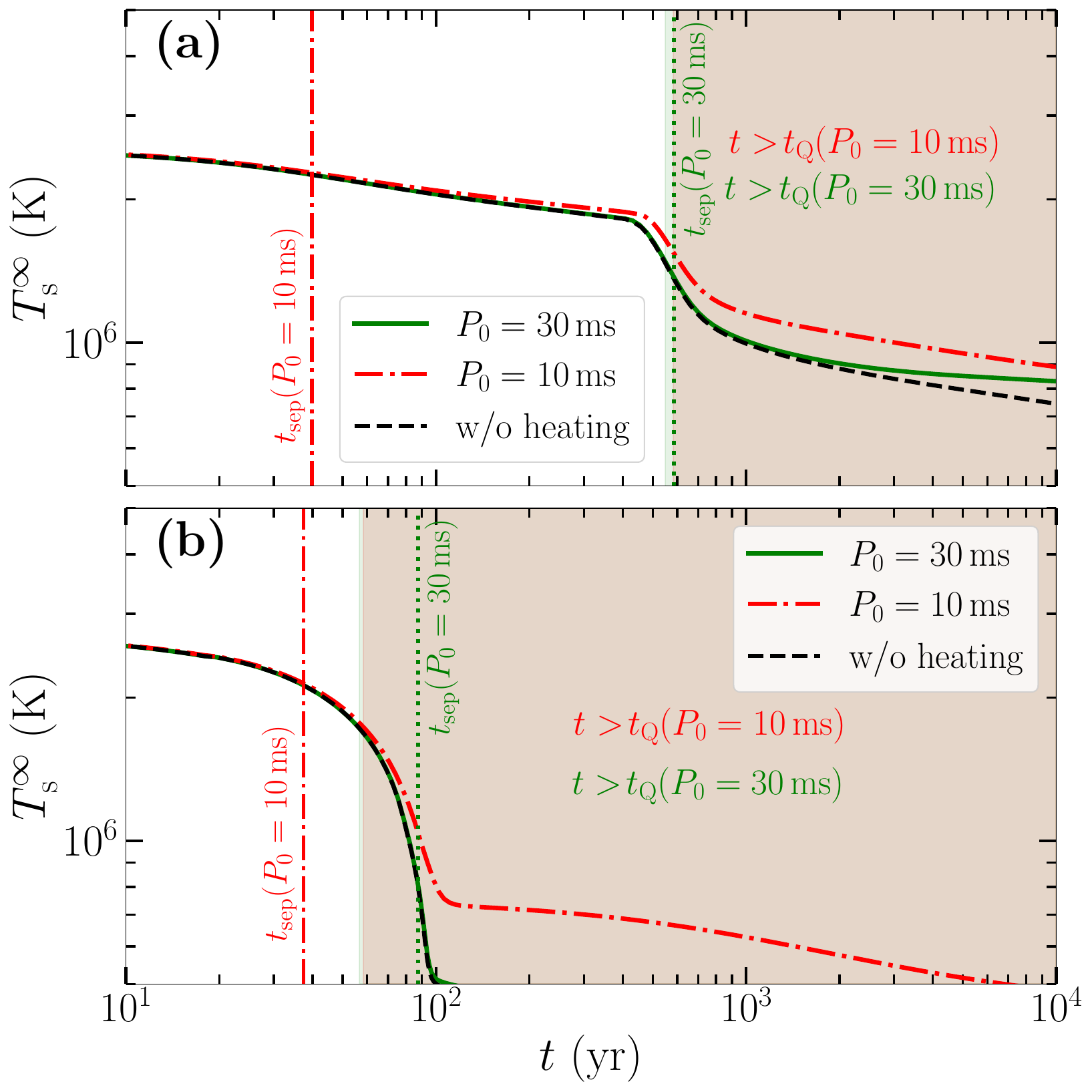}
\caption{Cooling curves for exemplary initial spin periods ($P_0=10$~ms and $30$~ms), illustrating where vortex creep heating begins to affect thermal evolution.
(a) $M=1.4\,M_\odot$ and (b) $M=2.0\,M_\odot$, both with $P\dot{P}=10^{-15}\,\mathrm{s}$.
Vertical lines mark the separation time $t_{\mathrm{sep}}$ (when heating produces $>1\%$ deviation) and the quantum-creep transition time $t_{\mathrm{Q}}$ (when $f_{\mathrm{Q}}>0.99$). Shaded region denotes epochs $t>t_{\mathrm{Q}}$ only; $t_{\mathrm{sep}}$ is treated separately and may lie either to the left or to the right of $t_{\mathrm{Q}}$.
A steady-state treatment is regarded as valid only when $t_{\mathrm{sep}}>t_{\mathrm{Q}}$.
Heavier stars enter the quantum-creep regime earlier due to direct Urca cooling.}
\label{quantum_creep_region_1.4M_and_2.0M}
\end{figure}

Figure~\ref{quantum_creep_region_1.4M_and_2.0M} shows representative cases for $M = 1.4\,M_\odot$ and $2.0\,M_\odot$ with $P\dot{P}=10^{-15}\,\mathrm{s}$. For smaller $P_0$, heating becomes significant too early ($t_{\mathrm{sep}} < t_{\mathrm{Q}}$), invalidating steady-state modeling, whereas larger $P_0$ satisfies $t_{\mathrm{sep}} > t_{\mathrm{Q}}$.

\begin{figure}[t]
\centering
\includegraphics[width=1.0\linewidth]{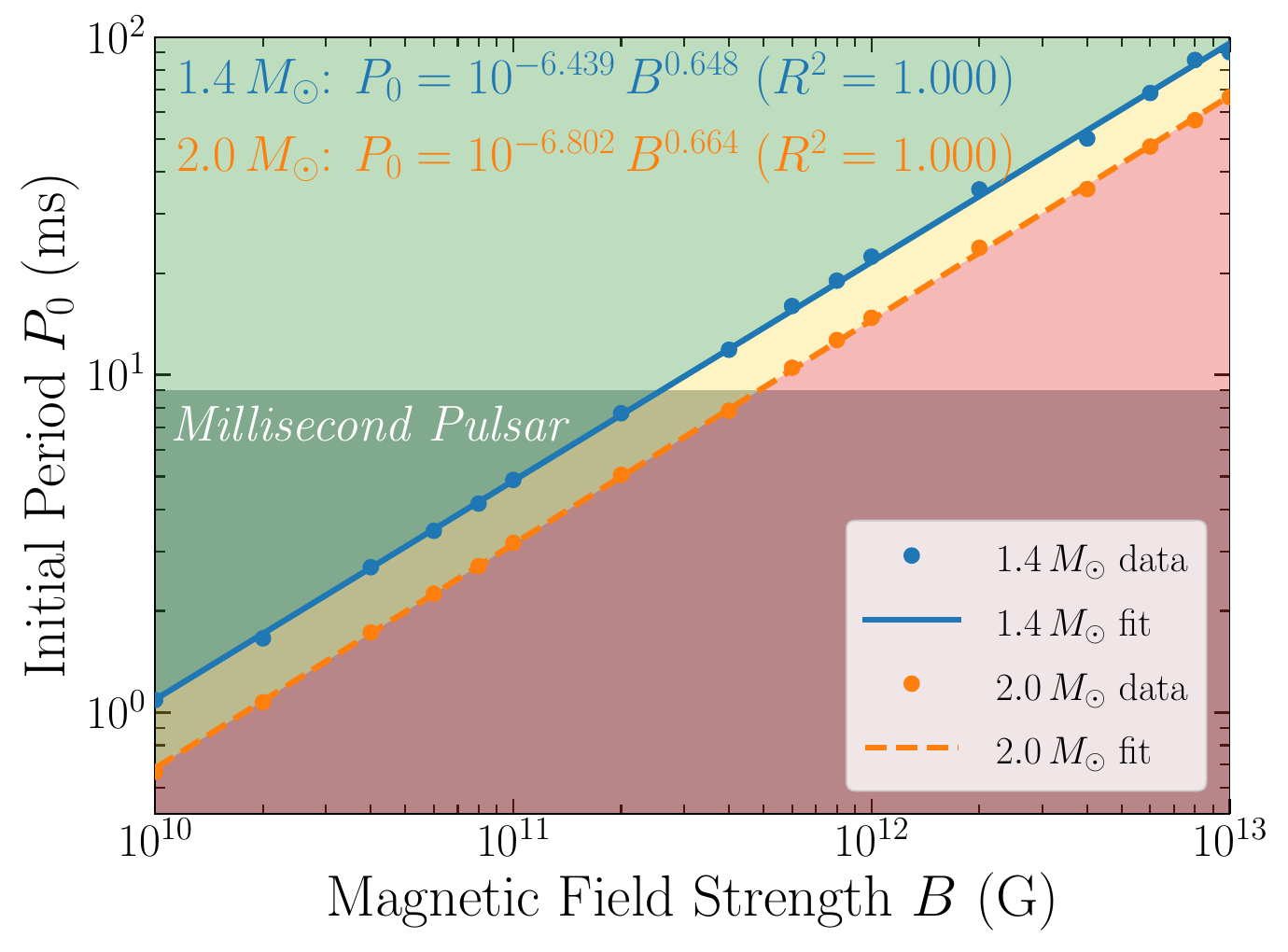}
\caption{Allowed region of $(B, P_0)$ satisfying $t_{\mathrm{sep}}>t_{\mathrm{Q}}$, where
$L_{\mathrm{h}} = J|\dot{\Omega}_\infty|$ remains valid without invoking additional
thermo--rotational dynamics.
Blue and orange boundary points correspond to $t_{\mathrm{sep}}\simeq t_{\mathrm{Q}}$ for 
$M=1.4\,M_\odot$ and $2.0\,M_\odot$, respectively, and follow the fitted relation 
$P_0 \propto B^{2/3}$, consistent with the analytic scaling 
$|\dot{\Omega}_\infty|\propto B^2/P_0^3$ under magnetic dipole spin-down.
The green shaded region marks parameter combinations that satisfy the validity condition
for both masses (allowed), while the red region indicates where the condition fails 
for either mass (not allowed). The intermediate yellow region represents parameter sets 
that are valid for one mass but not the other.
The gray band ($P_0\lesssim9$~ms) highlights the millisecond-pulsar regime, where rapid rotation
violates the spherical-symmetry assumptions of our cooling model.}
\label{allowed_range}
\end{figure}

Finally, Fig.~\ref{allowed_range} presents the resulting constraints in the $B$--$P_0$ parameter space for both
$M=1.4\,M_\odot$ and $2.0\,M_\odot$, indicating where the use of $L_{\mathrm{h}}=J|\dot{\Omega}_\infty|$ remains
valid. The boundary separating valid and invalid regions is defined by $t_{\mathrm{sep}}=t_{\mathrm{Q}}$: above this
boundary ($t_{\mathrm{sep}}>t_{\mathrm{Q}}$), VCH becomes influential only after the star has entered
the quantum--creep steady state, whereas below it ($t_{\mathrm{sep}}<t_{\mathrm{Q}}$), heating would take effect before
quantum tunneling dominates, invalidating the steady--state approximation.

The nearly linear shape of these boundaries can be understood analytically. At $t_{\mathrm{sep}}\simeq t_{\mathrm{Q}}$,
the effect of heating is still $\lesssim1\%$, so cooling curves are essentially identical to those without heating.
Therefore, the scaling along the boundary is governed by the pure dependence of the heating power on spin down:
$L_{\mathrm{h}}\propto |\dot{\Omega}_\infty|$. For magnetic dipole spin-down,
\begin{align}
|\dot{\Omega}_\infty|(t)
&=
\frac{2\pi}{P_0^3}
\left(\frac{B}{C}\right)^2
\left( 1 + \frac{t}{\tau} \right)^{-3/2},
\end{align}
where
\begin{align}
\tau = \frac{P_0^2}{2P\dot{P}},\;
C=3.2\times10^{19}~{\rm G}.
\end{align}
For ordinary pulsars, $t_{\mathrm{Q}}\ll\tau$ holds, so there holds
\begin{align}
|\dot{\Omega}_\infty| \propto \frac{B^2}{P_0^3}.
\end{align}
Requiring that $|\dot{\Omega}_\infty|$ remain constant along the validity boundary gives
\begin{align}
P_0 \propto B^{2/3},
\end{align}
fully consistent with the fitted exponents ($\simeq0.65$) in Fig.~\ref{allowed_range}. The slight offset from the ideal exponent is expected, as the boundary is obtained by the approximate condition $t_{\mathrm{sep}}\approx t_{\mathrm{Q}}$ rather than enforcing the strict equality $t_{\mathrm{sep}}=t_{\mathrm{Q}}$. Residual numerical truncation errors, together with the practical difficulty of identifying the exact equality computationally, naturally lead to the small discrepancy.
This agreement confirms that the constraints are determined by the relative timing of when VCH
becomes dynamically relevant compared to the onset of quantum-creep dominance.

Figure~\ref{allowed_range} also shows that the allowed region extends to smaller $P_0$ for $2.0\,M_\odot$
than for $1.4\,M_\odot$, reflecting the earlier transition to the quantum-creep regime enabled by rapid
DUrca cooling. Although not explicitly shown in Fig.~\ref{allowed_range}, we have confirmed that an
intermediate mass such as $1.8\,M_\odot$ falls within the yellow region located between the two boundaries,
demonstrating that the $P_0 \propto B^{2/3}$ relation is universal across stellar masses. Only the absolute
positioning of the boundary shifts slightly due to mass dependence in the cooling curves. We note that
increasing the relative-difference criterion above 1\% would introduce more significant contributions from
neutrino and photon cooling, potentially weakening the clean power-law scaling. In addition, different
envelope compositions may shift the absolute boundaries while retaining the same $B^{2/3}$ dependence.
Overall, the derived constraints provide practical guidance for selecting spin parameters that enable a
reliable application of VCH without resorting to full thermo-rotational dynamics.

\subsection{Effect of the initial spin period $P_0$ on cooling curves}
\label{Sec:P0_dependence}

In Sec.~\ref{Sec:verification}, we validated our implementation of VCH 
by reproducing the results of Ref.~\cite{Fujiwara_2024}, and we
established the allowed domain of $(B,P_0)$ in which the simplified prescription
$L_{\mathrm{h}} = J |\dot{\Omega}_\infty|$ remains applicable. However, the
previous study considered only a single representative choice of spin parameters
($P\dot P = 10^{-15}\,\mathrm{s}$ and $P_0 = 10$\,ms for ordinary pulsars), which does not reveal the
broader impact of $P_0$ on the thermal evolution. Since the spin-down rate is
\begin{align}
|\dot{\Omega}|(t) =
\frac{\pi}{\sqrt{2P\dot P}}
\!\left(t + \frac{P_0^2}{2P\dot{P}}\right)^{-3/2},
\label{eq:heating_spin_down_rate_reuse}
\end{align}
both the magnitude and the onset timing of VCH depend sensitively on the initial
spin. We note that $\dot{\Omega}$ denotes the magnetic dipole spin-down rate, whereas $\dot{\Omega}_\infty$ refers to the currently observed rate under the assumption of a steady state. In practice, however, the two can be used interchangeably for our purposes: VCH becomes thermally relevant only after the system has effectively reached the steady state, and even before that point its heating contribution remains negligible compared with the neutrino luminosity.

In this section, we explore the parameter region
$10\,\mathrm{ms} \le P_0 \le 570\,\mathrm{ms}$ motivated by the distribution of
initial periods $P_0$ inferred from Ref.~\cite{Igoshev_2022_initial_periods}, while
including the canonical value $P_0 = 10$\,ms used in
Ref.~\cite{Fujiwara_2024}. We consider $B = 10^{10}$, $10^{11}$, $10^{12}$, and $10^{13}$\,G, 
for both $1.4\,M_\odot$ and $2.0\,M_\odot$ neutron stars, and for the iron
envelope ($\Delta M = 0$) and the carbon envelope
($\Delta M = 10^{-8}\,M_\odot$). The proportional constant $J$ is varied within
$J \simeq 10^{42.9\text{--}43.8}\,$erg\,s, following
Ref.~\cite{Fujiwara_2024}. Observational data used for comparison are taken from
Table~\ref{observational_data_updated}. To obtain a comprehensive view of the
$P_0$ dependence, results outside the allowed region of
Fig.~\ref{allowed_range} are also displayed; such cases must be interpreted with
caution regarding the validity of the simplified heating prescription $L_{\mathrm{h}} = J |\dot{\Omega}_\infty|$.

\subsubsection{$1.4\,M_\odot$ neutron star: Iron envelope}

\begin{figure*}[t]
\centering
\includegraphics[width=\linewidth]{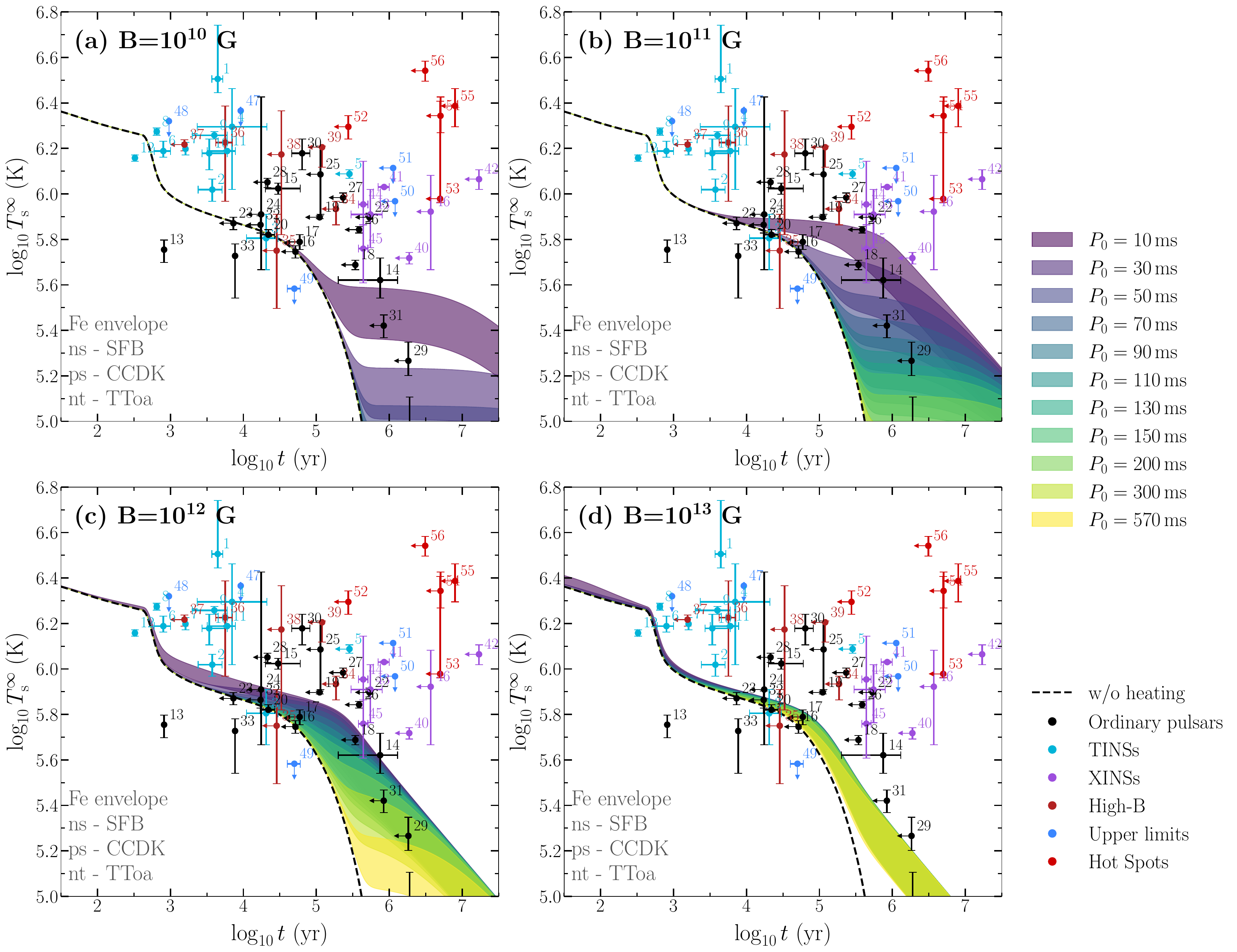}
\caption{
Cooling curves for a $1.4\,M_\odot$ neutron star with an iron envelope,
showing the dependence on $P_0$ for
(a) $B = 10^{10}$\,G, (b) $10^{11}$\,G,
(c) $10^{12}$\,G, and (d) $10^{13}$\,G.
Dark colors indicate smaller $P_0$.
Solid bands correspond to VCH with $J \simeq 10^{42.9\text{--}43.8}\,$erg\,s,
and dashed curves show cooling without VCH.
Observational points reflect the updated dataset in
Table~\ref{observational_data_updated}.}
\label{1.4M_iron}
\end{figure*}

Figure~\ref{1.4M_iron} shows that decreasing $P_0$ causes VCH to affect the
cooling curve earlier and more strongly, owing to the $P_0^{-3}$ dependence of
$|\dot{\Omega}(0)|$. For $B = 10^{10}$\,G [panel (a)], heating signatures appear
only when $T_{\mathrm{s}} \lesssim 10^{5}$\,K for $P_0 \gtrsim 70$\,ms,
implying that such weak fields make VCH observationally challenging. As $B$
increases to $10^{11}$\,G [panel (b)], heating becomes sufficiently strong to
impact the neutrino cooling era, allowing $P_0$ values up to $\sim200$\,ms to be distinguished 
in the regime where $T_{\mathrm{s}} \gtrsim 10^{5}$\,K. For $B = 10^{12}$\,G [panel (c)],
the heating effect substantially strengthens: for $P_0 = 10$\,ms, VCH becomes influential even before the rapid temperature
drop associated with the neutron $^3\mathrm{P}_2$ PBF process, whereas larger
$P_0$ values show progressively delayed onset consistent with the reduced
initial spin-down power. At $B = 10^{13}$\,G [panel (d)], VCH begins even
earlier but becomes insufficient to sustain high heating luminosity at late times
because $|\dot{\Omega}|$ decays to very small values; all curves eventually
converge toward a common $t^{-3/2}$ decline. Notably, the validity constraint
in Fig.~\ref{allowed_range} requires $P_0 \gtrsim \mathcal{O}(20$--$100$\,ms)
in such high-$B$ systems, indicating that some rapidly rotating cases in
panels (c)-(d) may fall outside the steady-state quantum-creep regime.

Importantly, the Cas~A~NS---whose rapid cooling and
its relation to neutron ${}^3\mathrm{P}_2$ gap models were the main focus of
our recent study \cite{Nam_2025_data_driven_cas_a_ns}---has not yet shown any
pulsations, and is inferred to possess a relatively weak magnetic field,
$B \lesssim 10^{11}\,\mathrm{G}$ \cite{Ho_2011_cas_a_magnetic}. As indicated by
its observational point (ID~12) in Fig.~\ref{1.4M_iron}(a)--(b), VCH is not
expected to noticeably influence Cas~A~NS at its current age even if
$P_0 = 10$\,ms, consistent with the omission of VCH in
Ref.~\cite{Nam_2025_data_driven_cas_a_ns}. Tighter observational constraints on
its surface magnetic flux density may allow a more precise prediction of when
VCH effects could emerge; however, given the currently inferred weak surface magnetic flux density, 
VCH is unlikely to be observationally confirmed in
Cas~A~NS in the foreseeable future.

To assess how the envelope composition influences the observable impact of VCH,
we now consider a light-element (carbon) envelope model.

\subsubsection{$1.4\,M_\odot$ neutron star: Carbon envelope}

\begin{figure*}[t]
\centering
\includegraphics[width=\linewidth]{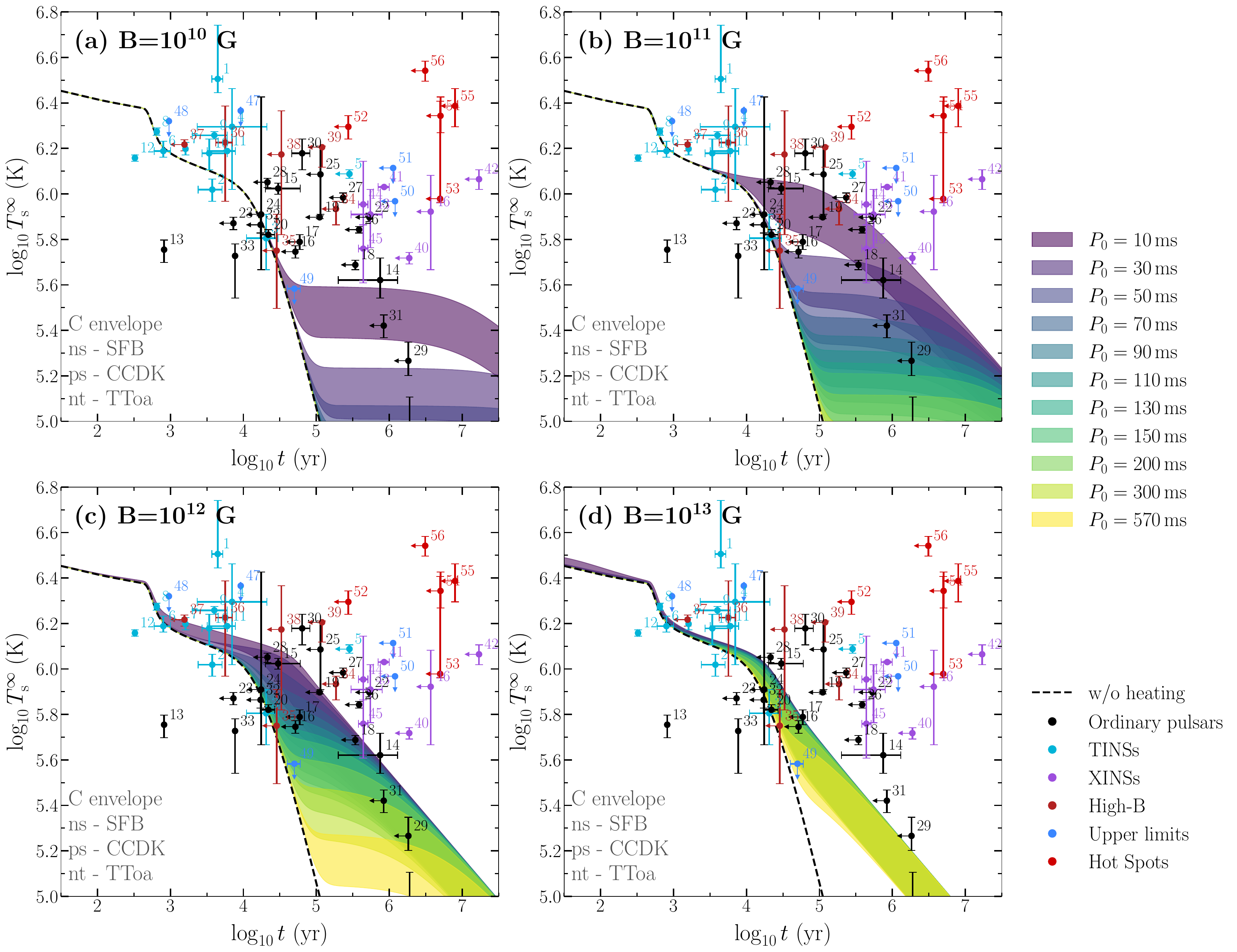}
\caption{
Same as Fig.~\ref{1.4M_iron}, but for a carbon envelope
($\Delta M = 10^{-8} M_\odot$).
The carbon envelope raises the surface temperature for a given internal
temperature and shift the onset of VCH impacts to earlier times.}
\label{1.4M_carbon}
\end{figure*}

Figure~\ref{1.4M_carbon} exhibits similar trends to the iron case, but with
higher surface temperatures and more rapid cooling. The earlier onset of VCH 
leads to an apparent lengthening of the plateau; however, 
caution is warranted because the logarithmic time axis
partially exaggerates this effect geometrically. 
Although VCH improves agreement with some warm sources such as TINSs and XINSs,
several others---including additional TINS/XINS objects, some High-B pulsars,
and the hottest hot-spot sources---remain unexplained, indicating the possible
need for supplementary heating mechanisms or alternative physical modeling.

In addition, sources with unusually low temperatures at early times (\textit{e.g.},
points ID~13 and ID~33) hint at the presence of fast neutrino emission channels such
as DUrca. We therefore turn to the $2.0\,M_\odot$ case below,
where DUrca cooling becomes relevant.

\subsubsection{$2.0\,M_\odot$ neutron star: Iron envelope}

Having established the $P_0$ dependence in the $1.4\,M_\odot$ case, we now
consider a more massive neutron star ($2.0\,M_\odot$), where the DUrca process operates. 
For the BSk24 EoS, the DUrca threshold lies at
$1.595\,M_\odot$, making the $2.0\,M_\odot$ configuration an ideal testbed for
the interplay between fast neutrino cooling and VCH. The
envelope composition, gap models, and heating prescriptions are identical to
those in the $1.4\,M_\odot$ analysis.

\begin{figure*}[t]
\centering
\includegraphics[width=\linewidth]{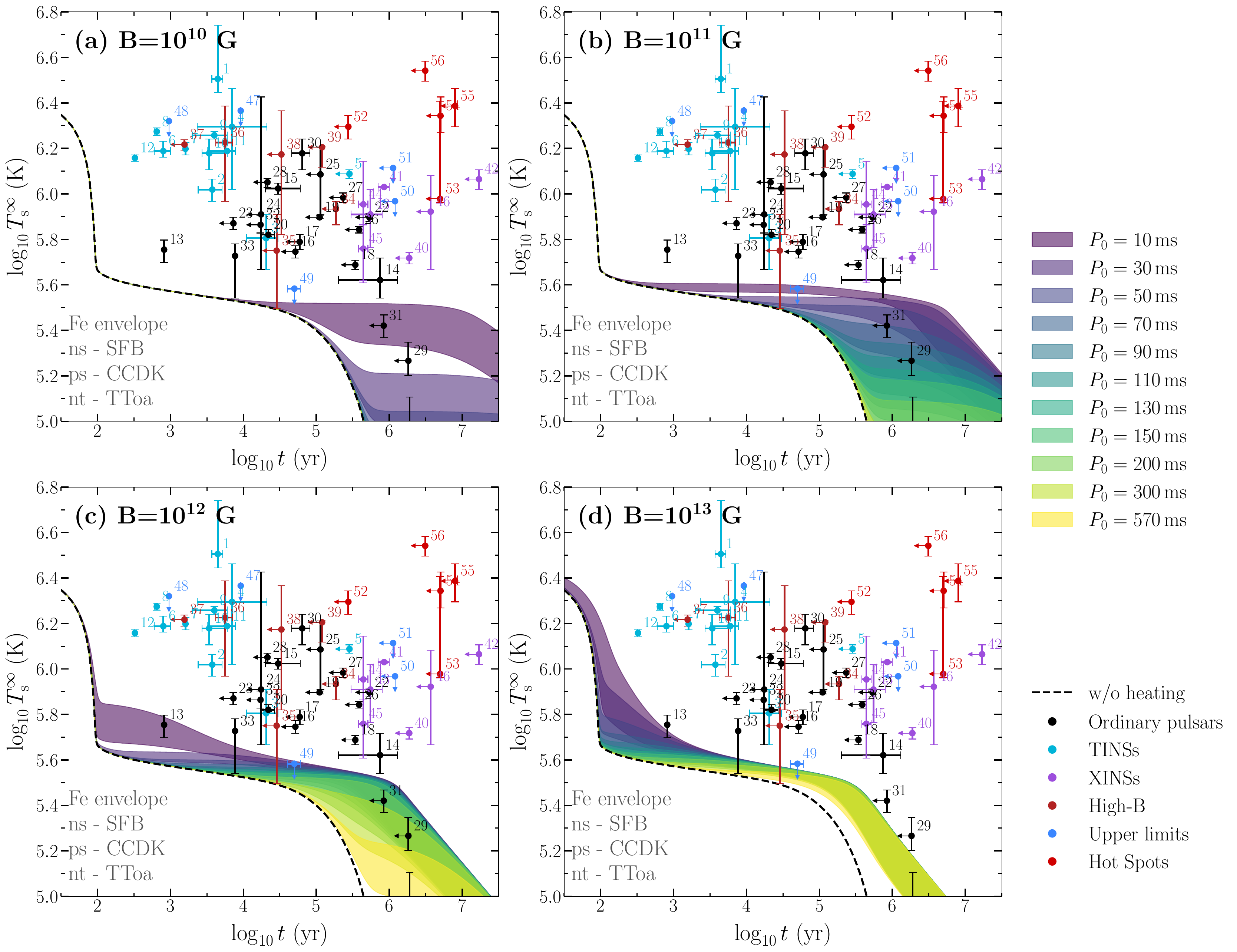}
\caption{
Cooling curves for a $2.0\,M_\odot$ neutron star with an iron envelope,
showing the dependence on $P_0$ for
(a) $B = 10^{10}$\,G, (b) $10^{11}$\,G,
(c) $10^{12}$\,G, and (d) $10^{13}$\,G.
DUrca cooling causes a rapid temperature drop at $t\sim10^2$\,yr.
VCH with $J \simeq 10^{42.9\text{--}43.8}\,$erg\,s produces
significant heating for sufficiently small $P_0$.}
\label{2.0M_iron}
\end{figure*}

Figure~\ref{2.0M_iron} reveals overall behaviors similar to the $1.4\,M_\odot$
case but strongly modified by the presence of DUrca cooling. For
$B = 10^{10}$\,G [panel (a)], VCH becomes visible only for
$P_0 \lesssim 50$\,ms and at $t \gtrsim 10^{4}$\,yr, since the weak magnetic
field yields a small $|\dot{\Omega}|$ and thus delayed heating relative to
the DUrca-induced rapid cooling. As $B$ increases to $10^{11}$\,G [panel (b)],
$P_0 = 10$\,ms already exhibits the heating effect by VCH immediately after the DUrca drop, 
and the heating that is distinguishable in the region of $T\gtrsim10^5$ K persists up to $P_0 \sim 150$--200\,ms.
For $B = 10^{12}$\,G [panel (c)], VCH becomes competitive even during the DUrca drop
for $P_0 = 10$\,ms, and heating signatures remain visible up to
$P_0 = 570$\,ms when $T_{\mathrm{s}} \gtrsim 10^5$\,K. For
$B = 10^{13}$\,G [panel (d)], the onset of VCH shifts to very early times, but
the late-time heating luminosity becomes insufficient due to the rapid decay of
$|\dot{\Omega}| \propto t^{-3/2}$.

It is crucial to emphasize that small $P_0$s in panels (c)--(d) may violate the
steady-state assumption underlying $L_{\mathrm{h}} = J|\dot{\Omega}_\infty|$.
According to the validity constraint in Fig.~\ref{allowed_range}, high-field
systems require $P_0 \gtrsim \mathcal{O}(15$--$70$\,ms) to ensure
$t_\mathrm{sep} > t_\mathrm{Q}$. Thus, configurations lying outside the allowed
domain must be interpreted with caution.

Notably, observational points ID~13, ID~33, and ID~49 can all be reproduced by the same evolutionary track with $B \ge 10^{12}\,$G when both DUrca cooling and VCH are active. In the absence of heating, a massive ($2.0\,M_\odot$) star undergoing DUrca would cool far more rapidly, leading to much lower surface temperatures at these ages. Their relatively high observed temperatures therefore suggest that if these objects are indeed massive, an additional heat source such as VCH is required to sustain them at the presently inferred surface temperatures. Likewise, ID~29 and ID~31 provide only upper bounds on their ages and thus remain compatible with such massive-star evolutionary pathways. We emphasize, however, that this interpretation is subject to model caveats to be discussed in Sec.~\ref{3d-extend}.

\subsubsection{$2.0\,M_\odot$ neutron star: Carbon envelope}

\begin{figure*}[t]
\centering
\includegraphics[width=\linewidth]{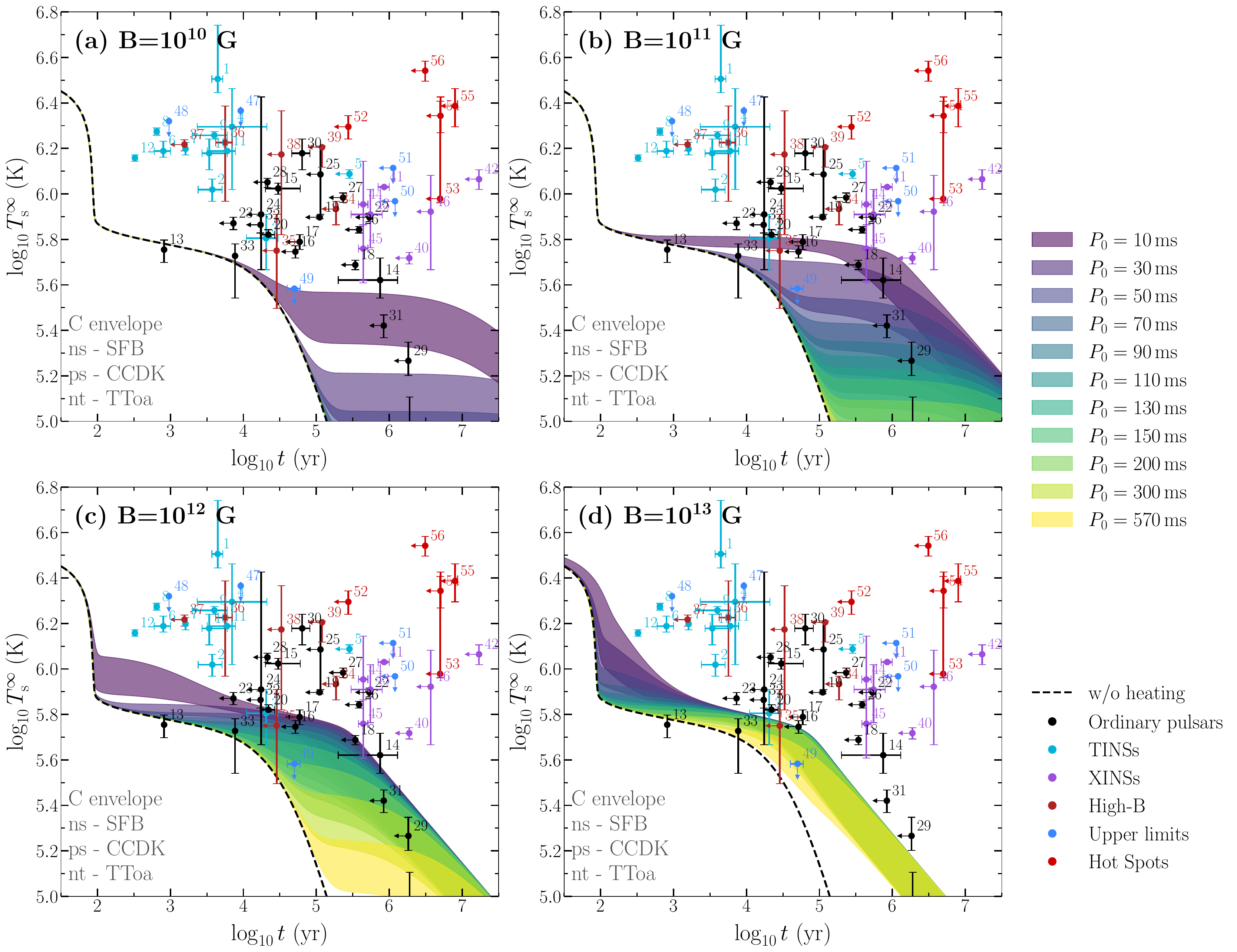}
\caption{
Same as Fig.~\ref{2.0M_iron}, but for a carbon envelope
($\Delta M = 10^{-8} M_\odot$).
The carbon envelope raises the surface temperature and shifts the observable
signatures of VCH to earlier times.}
\label{2.0M_carbon}
\end{figure*}

Figure~\ref{2.0M_carbon} shows comparable behavior to the iron-envelope case,
but with systematically higher $T_{\mathrm{s}}$ and more rapid early cooling,
owing to the higher thermal conductivity of the light-element envelope which
enhances heat transport to the surface. In particular, cooling-only models
can already reproduce ID~13 and ID~33 through DUrca cooling in a massive star
with a carbon envelope. The inclusion of VCH further broadens the accessible
parameter space: objects such as ID~29 and ID~31---difficult to reconcile under
cooling-only scenarios---can instead remain warm if VCH provides sustained
late-time heating.

Overall, these results demonstrate that VCH can play a 
significant role in shaping the observable thermal evolution of neutron stars, 
but only when the spin-down power $L_\mathrm{h}$ remains sufficiently large. 
The strong sensitivity of VCH signatures to the initial spin period $P_0$
(and, implicitly, to $P\dot{P}$) highlights that rotational parameters can play a
more decisive role than stellar mass in determining the late-time thermal state.
This emphasizes that $P_0$---a parameter often underexplored in previous cooling
studies---can critically influence whether a neutron star remains warm, even in
the presence of fast DUrca cooling in massive stars with sufficiently strong
magnetic fields.

Having established the strong two-parameter dependence on $(B, P_0)$ in the 
traditional $\log_{10} t$-$\log_{10} T^{\infty}_{\mathrm{s}}$ plane, we now 
extend our analysis into a three-dimensional (3D) parameter space to uncover 
systematic structures and population distributions shaped by surface magnetic flux density.
This higher-dimensional picture will also reveal interpretational 
caveats of the kind highlighted above, which we discuss further in Sec.~\ref{3d-extend}.

\subsection{Extending cooling curves into the 3D parameter space}
\label{3d-extend}

The conventional cooling representation, \textit{i.e.}, the 
$\log_{10} t$--$\log_{10} T^{\infty}_{\mathrm{s}}$ plane, hides an important 
dimension of the parameter space: neutron stars with vastly different magnetic 
fields $B$ may occupy almost identical positions when projected onto two 
dimensions. In particular, some ordinary pulsars with measured $(P,\dot{P})$ 
(black points) appear to trace nearly 
the same thermal evolution tracks in Figs.~\ref{1.4M_iron} 
and \ref{2.0M_iron}, despite possessing different magnetic fields. 
This motivates us to introduce $\log_{10} B$ as an additional axis to 
disentangle their true evolutionary pathways.

\begin{figure*}[t]
\centering
\includegraphics[width=\linewidth]{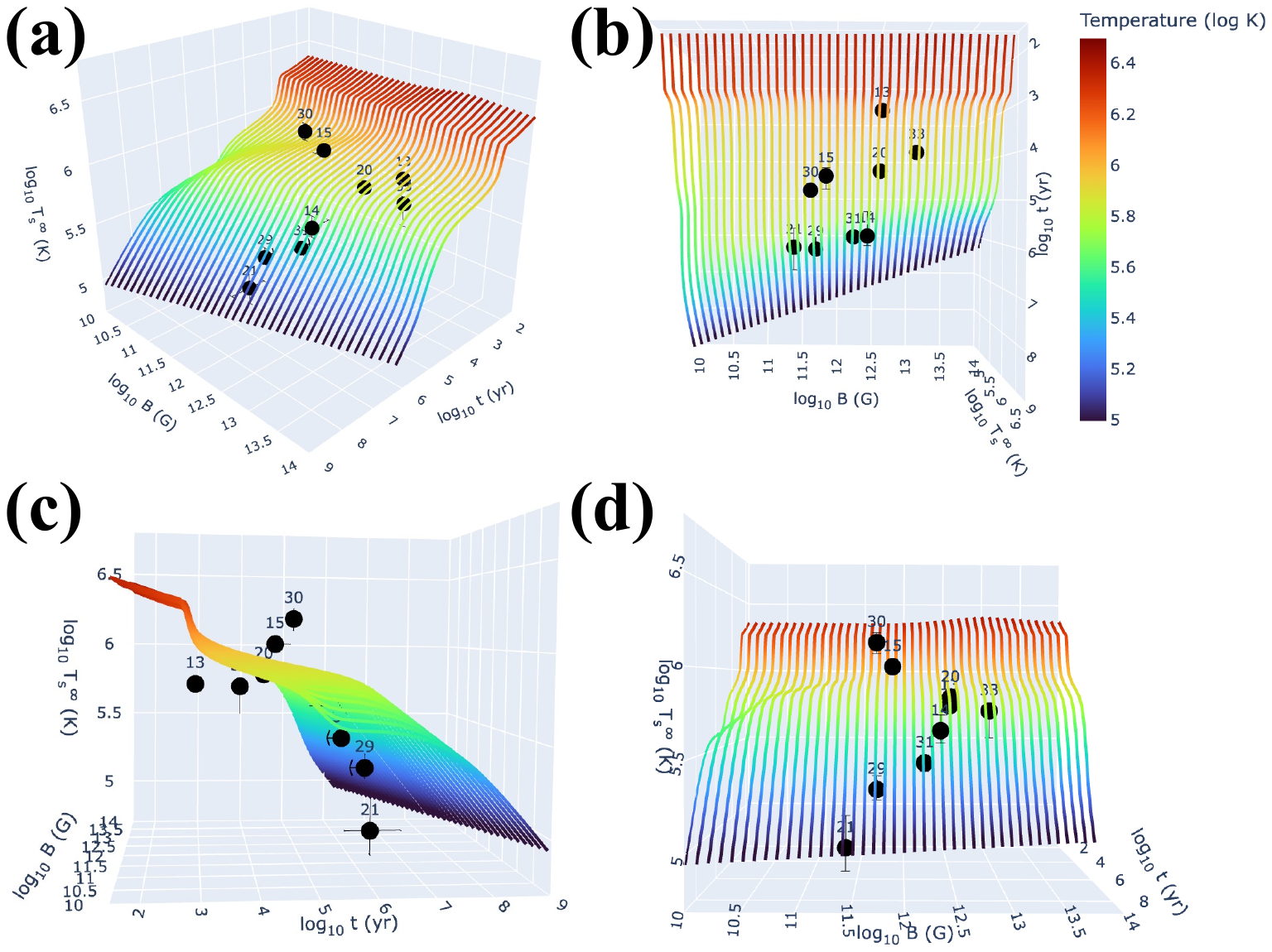}\vspace{-3mm}
\caption{
Three-dimensional extension of the cooling curves for a $1.4\,M_{\odot}$ neutron star 
with $P_0 = 10$\,ms, $J=10^{43.8}\,\mathrm{erg\,s}$, and an iron envelope. 
(a) Full perspective view of the cooling surface and ordinary pulsar sample. 
(b) Top-down view demonstrating magnetically stratified evolutionary pathways 
and confirming the expected scaling $t \propto B^{-2/3}$ at 
$T_{\mathrm{s}}^{\infty}\sim10^{5}$\,K. 
(c) Projection along the $\log_{10} B$ axis illustrating the extension of the 
conventional cooling curves into the magnetic-field dimension. 
(d) Projection compressing the time axis, highlighting clustering in the 
$(\log_{10} B,\log_{10} T_{\mathrm{s}}^{\infty})$ plane. 
Surface magnetic flux density is sampled over $\log_{10} B = 10.0,\,10.1,\,\dots,\,13.9,\,14.0$ to ensure 
smooth representation of the cooling surface.
Only ordinary pulsars with measured $(P,\dot{P})$ in 
Table~\ref{observational_data_updated} are shown.
}
\label{3d_plots}
\end{figure*}

Figure~\ref{3d_plots} presents a 3D extension of the cooling 
curves for a $1.4\,M_{\odot}$ neutron star with 
$P_0=10$\,ms, $J=10^{43.8}$\,erg\,s, and an iron envelope.
Panel (a) provides a perspective view, displaying both the thermal-evolution 
surface and the distribution of ordinary pulsars in $(t,B,T_{\mathrm{s}}^{\infty})$ space.
Panel (b) compresses the $T_{\mathrm{s}}^{\infty}$ axis, revealing a clear
vertical stratification by the surface magnetic flux density: sources that are apparently
coincident in the $\log_{10} t$--$\log_{10} T_{\mathrm{s}}^{\infty}$ plane
separate into distinct $B$ layers once the third dimension is included.
Panel (c) rotates the view such that the $\log_{10} B$ axis is nearly aligned 
with the line of sight, demonstrating how the conventional cooling curves 
extend smoothly into the magnetic-field dimension.
Panel (d) highlights relative positions in the 
$(\log_{10} B,\log_{10} T_{\mathrm{s}}^{\infty})$ projection, offering additional 
discriminatory power for sources with similar age.

A physically consistent trend becomes evident in panel~(b): the locus where 
$T_{\mathrm{s}}^{\infty}=10^{5}$\,K follows a power law,
\begin{align}
\log_{10} t \simeq -0.663\,\log_{10} B + 15.55,
\end{align}
which directly results from the spin-down dependence 
$|\dot{\Omega}|\propto t^{-3/2}$ at late times 
and the balance $L_{\mathrm{h}}\simeq L_{\gamma}$.
In the late-time limit $t\gg\frac{P_0^2}{2P\dot{P}}$,
$|\dot{\Omega}|\approx \dfrac{\pi}{\sqrt{2P\dot{P}}}\,t^{-3/2}$ 
with $P\dot{P}\propto B^{2}$; equating 
$L_{\mathrm{h}} = J|\dot{\Omega}|$ to the fixed 
$L_{\gamma}(T_{\mathrm{s}}^{\infty}=10^{5}\,\mathrm{K})$ 
gives $B^{-1} t^{-3/2}=\mathrm{const.}$, 
and hence $t\propto B^{-2/3}$. 
This agreement confirms that our VCH implementation 
recovers the theoretically expected scaling, 
reinforcing the physical validity of our numerical treatment.

Taken together, these visualizations emphasize an important caveat for interpreting 
two-dimensional cooling plots. In particular, as discussed in 
Sec.~\ref{Sec:P0_dependence}, sources ID~13, ID~33, ID~29, and 
ID~31 appear to follow similar thermal evolutionary pathways in 
Fig.~\ref{2.0M_iron}, but their distinct magnetic fields place them on 
separate evolutionary sheets of the cooling surface. Thus, even if two sources 
share the same $(t,T_{\mathrm{s}}^{\infty})$, their spin and magnetic histories 
may be fundamentally different. Therefore, such objects should not be assumed to share a common evolutionary
track based solely on their positions in the two-dimensional cooling plane.

Overall, these results demonstrate that when VCH contributes to the heating 
source, magnetic field must be treated as a genuinely independent axis in the 
cooling analyses. We therefore recommend incorporating 3D evolutionary 
representations in future studies of isolated neutron-star observational data, 
particularly when interpreting sources near the warm tail where rotational 
heating competes with fast neutrino cooling.

\section{Summary and prospect}
\label{Sec:Conclusion}

In this work, we have further extended our newly developed neutron-star cooling
framework~\cite{Nam_2025_data_driven_cas_a_ns} to incorporate vortex creep heating (VCH),
building upon the first results reported in our INPC2025 proceedings
contribution~\cite{Nam_2025_INPC}. While most previous VCH studies focused on
relatively light neutron stars and late evolutionary stages
($t \gtrsim 10^{5}$\,yr), the combined action of VCH and fast direct Urca (DUrca) neutrino cooling in massive stars has remained largely unexplored. Using the
BSk24 equation of state, for which DUrca activates above $M \approx 1.6\,M_\odot$,
we have performed the first systematic investigation of VCH in a
$2.0\,M_\odot$ neutron star undergoing rapid DUrca cooling.

A key advancement of this study is the introduction of the quantum-creep coverage
fraction $f_{\mathrm{Q}}$ (Eq.~(\ref{def_f_Q})), which quantitatively tracks whether the vortex creep
process has entered the steady-state quantum-tunneling regime. This allowed us
to define a physically consistent validity domain in the $(B,P_0)$ plane where
the simplified prescription $L_{\mathrm{h}}=J|\dot{\Omega}_\infty|$ remains
reliable. In particular, we have demonstrated that for sufficiently high magnetic
fields, the onset of VCH can occur too early---before the entire inner crust has
fully cooled into the quantum-creep regime---and such cases must be treated with
caution. We have therefore proposed a practical criterion
(Fig.~\ref{allowed_range}) to evaluate the physical consistency of VCH heating
calculations.

We have further showed that the fraction of superfluid moment of inertia $\chi_{\mathrm{sf}}$ [Eq.~(\ref{def_chi_sf})] remains large
in both $1.4\,M_\odot$ and $2.0\,M_\odot$ stars by the time heating becomes
important. This suppresses the dependence of VCH heating on the neutron
${}^1\mathrm{S}_0$ gap model, justifying the use of standard gap choices without strongly
affecting observational predictions.

Our results have revealed a possibility that previously puzzling \emph{old warm} neutron stars could be far more massive than previously thought, whose DUrca-induced rapid cooling is partially
compensated by VCH. We demonstrated that both iron and light-element (carbon)
envelopes permit such warm late-time solutions for reasonable combinations of
$(B,P_0)$, providing a compelling alternative interpretation of several sources
with high temperatures at ages $t \gtrsim 10^5$ yr.

Finally, we have introduced a three-dimensional extension of cooling curves into the
$\log_{10} t$--$\log_{10} T^{\infty}_{\mathrm{s}}$--$\log_{10} B$ space
(Fig.~\ref{3d_plots}). This visualization reveals that $(t,T_{\mathrm{s}}^{\infty})$
alone is insufficient to uniquely determine the evolutionary state of a neutron
star when VCH is active: sources appearing
coincident in the traditional cooling plane may occupy distinct magnetic-field
layers. The presence of physically consistent scaling relations such as
$t\propto B^{-2/3}$ further validates our VCH implementation and highlights the
crucial role of $B$ in shaping thermal evolution.

Looking ahead, we will extend the present framework to highly magnetized neutron
stars by developing a two-dimensional cooling solver including anisotropic heat
transport and field evolution. We also aim to calibrate the proportional
constant $J$ through microscopic modeling of vortex pinning forces across
multiple equations of state (\textit{e.g.}, Refs.~\cite{Wlazlowski(2016),Klausner(2023)}),
and to incorporate additional internal heating mechanisms such as rotochemical
heating~\cite{Reisenegger_1995_rotochemical_1,Fernandez_2005_rotochemical_2,%
Petrovich_2010_rotochemical_3}, magnetic-field decay~\cite{Pons_2009_magnetic_field_decay_1,%
Vigano_2013_magnetic_field_decay_2}, and dark-matter heating
(e.g., Refs.~\cite{Kouvaris_2008_dark_matter_1,Lavallaz_2010_dark_matter_2,%
Baryakhtar_2017_dark_matter_3}).

In parallel, following the data-driven strategy introduced in our previous work
on the Cassiopeia A neutron star~\cite{Nam_2025_data_driven_cas_a_ns}, we aim to explore the parameter
space spanned by $(P\dot{P},\,P_0,\,M,\,\Delta M)$ directly constrained by
current available observational data. This approach will allow us to identify
which regions of the parameter space remain observationally viable for
VCH-active tracks and will lay the groundwork for a Bayesian inference
framework that can ultimately determine confidence intervals for such
evolutionary pathways.

\section*{Acknowledgments}

This work is supported by JSPS Grant-in-Aid for Scientific Research, Grants No.~23K03410, No.~23K25864, and No.~JP25H01269.

\appendix

\section{$P$--$\dot{P}$ diagram for the observational sample}
\label{Appendix:PPdot}

\begin{figure*}[t]
  \centering
  \includegraphics[width=\linewidth]{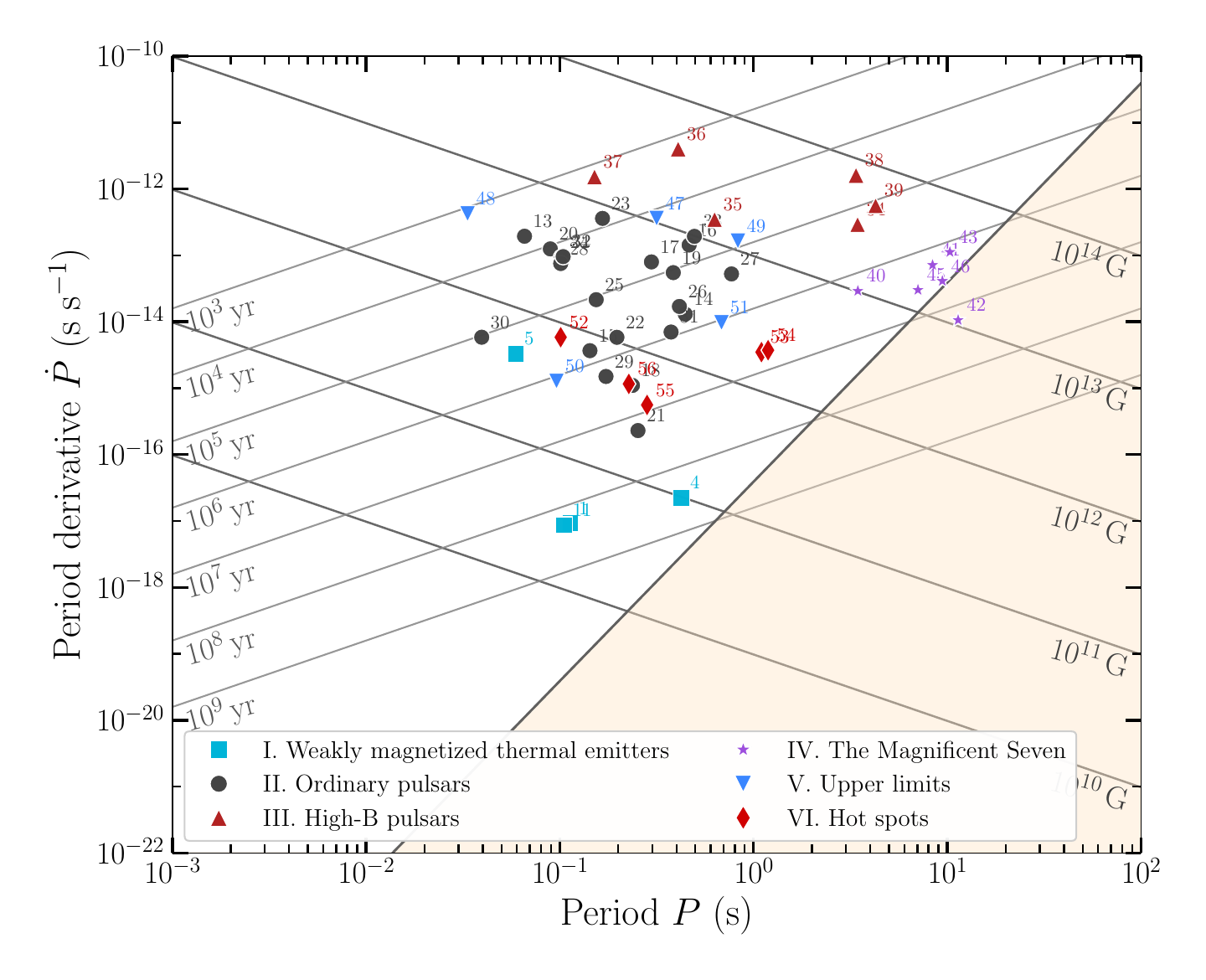}\vspace{-7mm}
  \caption{
  $P$--$\dot{P}$ diagram for the observational samples listed in 
  Table~\ref{observational_data_updated}. 
  Spin period and period derivative are used to estimate the characteristic age 
  (diagonal upward bands) and surface magnetic flux density assuming magnetic 
  dipole braking (diagonal downward bands). 
  The shaded lower-right region denotes the radio death zone. 
  The Magnificent Seven exhibit strong magnetic fields and lie near the death 
  line, while ordinary pulsars span 
  $B \sim 10^{11\text{--}13}\,$G across a wide age range. 
  Only sources with measured $(P,\dot{P})$ are plotted.
  }
  \label{P_Pdot_diagram}
\end{figure*}

Figure~\ref{P_Pdot_diagram} shows the $P$--$\dot{P}$ distribution of the 
observational samples in Table~\ref{observational_data_updated}, restricted to 
sources for which both spin period $P$ and period derivative $\dot{P}$ are available. 
Different classes are marked with distinct colors and symbols for clarity.

The diagonal bands rising toward the upper right indicate characteristic ages
$\tau_{\mathrm{c}} \equiv P/(2\dot{P})$, while the bands falling toward the 
lower right denote magnetic dipole estimates of the surface magnetic flux 
density, $B \propto \sqrt{P\dot{P}}$. The shaded region in the lower-right 
portion of the diagram represents the “death zone,” where radio emission is no 
longer expected to be sustained due to insufficient pair production.

As expected, the Magnificent Seven occupy the region of 
$B \sim 10^{13\text{--}14}\,$G and lie close to the death line, consistent with 
their radio silence and thermal X-ray emission. Ordinary pulsars---the primary 
targets of this work---are broadly distributed across 
$B \sim 10^{11\text{--}13}\,$G and 
$\tau_{\mathrm{c}} \sim 10^{4\text{--}7}\,$yr. 
Weakly magnetized thermal emitters cluster at lower magnetic fields 
($B \lesssim 10^{11}$\,G), with some members (e.g., ID~1, ID~4, ID~11) 
exhibiting large characteristic ages of $\gtrsim10^{7}$\,yr.
High-$B$ pulsars, while having magnetic fields comparable to the Magnificent 
Seven, tend to possess younger characteristic ages.

Sources classified as hot-spot objects or those with upper limits on 
$T_{\mathrm{s}}^{\infty}$ are intermixed with ordinary pulsars in this 
parameter space, showing no distinctive separation in terms of $(P,\dot{P})$ 
alone. This reinforces the necessity of combining spin parameters with 
thermal information---as done in this work---to elucidate their underlying 
evolutionary mechanisms.

\bibliography{ref}
\bibliographystyle{unsrt}

\end{document}